\documentclass[]{aa}

\usepackage{txfonts}
\usepackage{graphicx}

\begin{document}

\title{High-resolution mapping of interstellar clouds with near-infrared
scattered light}

\titlerunning{High-resolution mapping of interstellar clouds}

\author{
Mika Juvela \inst{1} \and
Veli-Matti Pelkonen \inst{1} \and
Paolo Padoan \inst{2} \and
Kalevi Mattila \inst{1}
}

\institute{
Helsinki University Observatory, FIN-00014, University of Helsinki, Finland
\and
Department of Physics, University of California, San Diego, 
CASS/UCSD 0424, 9500 Gilman Drive, La Jolla, CA 92093-0424; ppadoan@ucsd.edu
}

\date{Received 1 January 2005 / Accepted 2 January 2005}

\abstract
{With current wide-field near-infrared (NIR) instruments the scattered light
in the near-infrared can be mapped over large areas. Below $A_{\rm V}\sim
10$\,mag the surface brightness is directly proportional to the column
density, and at slightly higher column densities the saturation of the
intensity values can be corrected using the ratios of the intensity in
different NIR bands. Therefore, NIR scattered light provides a promising new
method for the mapping of quiescent interstellar clouds.}
{We develop a method to convert the observed near-infrared surface brightness
into estimates of the column density. We study and quantify the effect
that different error sources could have on the accuracy of such estimates.
We also propose to reduce systematic errors by combining surface brightness
data with extinction measurements derived from the near-infrared colour excess
of background stars.}
{Our study is based on a set of three-dimensional magnetohydrodynamic
turbulence simulations. Maps of near-infrared scattered light are obtained 
with radiative transfer calculations, and the maps are converted back into 
column density estimates using the proposed method. The results are compared 
with the true column densities. Extinction measurements are simulated using 
the same turbulence simulations, and are used as a complementary column density 
tracer.}
{We find that NIR intensities can be converted into a reliable estimate of 
the column density in regions with $A_{\rm V}$ up to almost 20\,mag. We show 
that the errors can be further reduced with detailed radiative transfer modelling 
and especially by using the lower resolution information available through the 
colour excess data.}
{We urge the observers to try this new method out in practice.}

\keywords{ISM: Structure -- ISM: Clouds -- Infrared: ISM -- dust, extinction -- Scattering 
-- Techniques: photometric }

\maketitle

\section{Introduction}

The large scale distribution of the interstellar matter is affected by
Galactic rotation, gravitational instabilities, and supernova explosions.
Within the clouds the structure results largely from supersonic turbulence
that is constantly fed by the large scaled phenomena and by stellar winds and
outflows from newborn stars. At smaller scales the self-gravity becomes the
dominant factor, keeping dense cloud cores together and eventually leading to
core collapse and the formation of new stars. The magnetic fields are
important at all scales. The flow of the matter is affected by a number of
processes, each having its own imprint on the observed density distributions.
Investigation of the cloud structure is important since it provides clues on
the relative importance of turbulence, self-gravity and magnetic fields at
different scales. It also provides information on different stages of the
star-formation process and sets constraints on theories of star formation.

Information about the 3D spatial structure of interstellar clouds is
obtained indirectly, through analysis of the observed radiation which
represents an integral along the whole line of sight. Extended maps of
interstellar clouds can be obtained with several methods: (1) the integrated
intensity of emission lines of various molecular or atomic tracers, especially
CO and HI, (2) the thermal emission of dust grains at far--IR and sub-mm
wavelengths, (3) star counts at optical or near-infrared wavelengths, (3) the
near-infrared reddening of the light from background stars, (4) mid-infrared
absorption toward dark clouds (Egan et al. \cite{Egan1998}; Hennebelle et al.
\cite{Hennebelle2001}), and (5) absorption toward bright x-ray background.
Each of these methods has limitations, but gives a complementary view of the
cloud structure. 

A given molecular line is useful only in a limited density range,
above the critical density of the transition, and below the optical depth
where the line saturates. Molecular abundances depend on complicated, time-dependent
chemical networks and, furthermore, depletion onto dust grains may cause
additional effects. In the study of molecular lines, one should be able to 
estimate the excitation of the molecules and possible radiative transfer
effects. Since the line-of-sight structure of the cloud is unknown, this cannot
be done accurately. Abundances and excitation vary both on the plane of the sky
and along each line of sight. It is well known that different molecules may
peak at entirely different locations and, therefore, the estimated column
density map depends critically on the selected tracer. Low extinctions around
$A_{\rm V}\sim$1 pose a special problem because of the transition from atomic
to molecular gas. Therefore, column density must be estimated by combining
information from several tracers with largely unknown abundances. The
resolution obtained by single dish observations is usually some tens of arc
seconds. Arc second resolution can be reached only with interferometric
observations of small areas.

Thermal dust emission at far-IR and sub-mm wavelengths could provide a more
straightforward picture of column density -- if dust temperatures can be
estimated and the dust to gas ratio remains constant. However, both dust
temperature and FIR/sub-mm optical properties of dust grains may have
significant spatial variations (Cambr\'esy et al. \cite{cambresy2001}, del
Burgo et al. \cite{delburgo2003}, Dupac et al. \cite{dupac2003}; Kramer et al.
\cite{kramer2003}, Stepnik et al. \cite{stepnik2003}; Lehtinen et al.
\cite{Lehtinen2004}, \cite{Lehtinen2006}; Ridderstad et al.
\cite{Ridderstad2006}). These may be due to grain growth by coagulation and
ice mantle deposition (Ossenkopf \& Henning \cite{ossenkopf1994}, Krugel \&
Siebenmorgen \cite{krugel1994}) or physical changes in the grain material
(e.g., Mennella et al. \cite{mennella98}, Boudet et al. \cite{Boudet2005}).
Quiescent clouds have rather low intensity in the FIR/sub-mm and, because of
the limited sensitivity, observations have concentrated on regions with
$A_{\rm V}\sim10^{\rm m}$ or above. For single dish observations the spatial
resolution is typically worse than $\sim$10$\arcsec$ in the sub-mm. FIR
observations must be carried out with space borne instruments and, in spite of
the shorter wavelengths, the resolution is not better.

Optical star counts are used to map the extinction mainly at low column
densities (Wolf \cite{wolf1923}) but can give reliable estimates up to $A_{\rm
V}\sim$5\,mag, depending on the available observations. High resolution is,
however, hard to obtain since each resolution element must contain a large
number of stars. In the optical region the stellar density drops rapidly above
$A_{\rm V}\sim$1, but in the NIR the star counting method remains useful
beyond $A_{\rm V}\sim$20, provided that deep K-band observations are
available. However, in NIR the colour excess method yields better spatial
resolution (e.g., Alves, Lada \& Lada \cite{alves2001}; \cite{cambresy2002}). 
The method is statistical in the sense that it relies on average properties of the
background stars. Unlike star counting, the colour excess method provides an
independent extinction estimate for each star, that is, for a number of very
narrow beams through the cloud. The errors of these individual extinction
estimates are dominated by the uncertainty on the (generally unknown) spectral
type of the background star. An actual map of extinction is obtained by
spatial averaging, and the reliability can be improved by combining results
from more than two NIR bands (Lombardi \& Alves \cite{lombardi2001}) and
adaptive spatial resolution (\cite{cambresy2002}). With
dedicated observations one can reach a resolution of $\sim$10$\arcsec$. In the
case of the commonly used 2MASS survey (limiting K$_{\rm s}$ magnitude
$\sim$15) the spatial resolution is, depending on the field location, a few
arc minutes and the covered extinction range $A_{\rm V}\sim$1--15\,mag.

The scattered light in dark clouds was first detected by photographic methods
at optical wavelengths by Struve and Elvey (\cite{Struve1936}) and Struve
(\cite{Struve1936}). They analyzed the light in terms of the dust scattering
properties. Later on, Mattila (\cite{Mattila1970a}a, \cite{Mattila1970b}b)
performed photoelectrical surface brightness observations of two dark nebulae
and determined the dust albedo and scattering asymmetry parameter at the UBV
bands. Haikala et al. (\cite{Haikala1995}) made the first imaging of scattered
light in a discrete diffuse/translucent cloud at the far UV wavelengths.

The first detection of NIR scattered light in a dark nebula illuminated by the
normal interstellar radiation field (ISRF) was reported by Lehtinen and
Mattila (\cite{Lehtinen1996}). They also studied the relationship between J,
H, and K band surface brightness vs. dust column density (measured by NIR
colour excess method) which they found to be linear up to optical depth of $\sim$1
in the wavelength band in question (their Fig. 8). From their Monte Carlo
light scattering calculations they predicted that the linear relationship will
saturate at optical depths of $\sim$1.5 to 2 and then turn down. They also made the
first determination of the albedo of Galactic interstellar grains in the JHK bands. The
idea of using NIR surface brightness as a high-resolution probe of the dust
density distribution in optically opaque clouds was presented by Lehtinen et
al. in the ESO Press release
26a/2003\footnote{http://www.eso.org/outreach/press-rel/pr-2003/phot-26-03.html}
in connection with their VLT/ISAAC observations of DC303.8-14.2.
Nakajima et al. (\cite{Nakajima2003}) presented JHK surface
brightness images of the Lupus 3 dark cloud. They discussed their measurements
in terms of scattered light from dust and presented diagrams of JHK surface
brightness vs. A$_{\rm V}$.

Padoan, Juvela, \& Pelkonen \cite{Padoan2006a} proposed scattered
near-infrared light as a direct measure of the cloud column density. That
study was motivated by the images of the Perseus region obtained by Foster \&
Goodman (\cite{Foster2006}), which, in accordance with Lehtinen \& Mattila
(\cite{Lehtinen1996}) and Nakajima et al. (\cite{Nakajima2003}), illustrated
that large scale mapping of scattered intensity has become possible even for
clouds illuminated by normal ISRF.
In the near-infrared the dust properties are believed to be rather constant,
resulting in smaller uncertainties in the scattering properties than at
shorter wavelengths. In the normal extinction curve (Cardelli
\cite{Cardelli1989}) the optical depth in the K-band is about one tenth of the
corresponding value in V-band. Therefore, in regions with $A_{\rm V}$ below
10\,mag, the K-band intensity remains almost directly proportional to the dust
column density. At higher extinctions the surface brightness starts to
saturate, the effect being stronger at shorter wavelengths. Padoan et al.
(\cite{Padoan2006a}) showed that if the saturation is taken into account, the
combination of J-, H-, and K-bands can be used to estimate column densities
for regions with $A_{\rm V}$ up to $\sim$20\,mag. The method avoids many
shortcomings of the other methods listed above. Most importantly, it provides
column density maps of interstellar clouds at a sub-arcsecond resolution. At
low $A_{\rm}$ the method is limited mainly by the sensitivity of the NIR
observations. 

In this paper we study in more detail the properties of column density
estimates that are based on near-infrared scattering. We use the method
presented by Padoan et al. (\cite{Padoan2006a}) as our starting point. We
examine and quantify the effects of possible error sources, including
differences in the cloud structure and column density, dust properties, 
and radiation field. We also address possible complications arising from
near-infrared dust emission and diffuse background surface brightness.
Furthermore, we propose to improve the reliability of the column density 
estimates with the help of colour excess measurements of background stars
(the reddening data are a byproduct of the observations) and address 
the possibility of more detailed radiative transfer modelling.

In Sect.~\ref{sect:method} we present the general method of converting NIR
surface brightness into column density. In Sect.~\ref{sect:modelling} we
present the turbulence and the radiative transfer calculations used in our
tests. The radiative transfer calculations provide simulated maps of
near-infrared surface brightness that can be converted back to column density
estimates using the proposed method. The results are compared with the true
column densities of our model clouds in Sect.~\ref{sect:tests}, where we will
also examine the possibility of using colour excess data of the background
stars to improve the accuracy of the column density estimates. In
Sect.~\ref{sect:discussion} we summarize our conclusions on the accuracy of
the proposed method and discuss some particular sources of error. Direct
radiative transfer modelling of observations is discussed in
Appendix~\ref{sect:detailed_modelling} and further discussion on some of the
error sources is provided in Appendix~\ref{sect:validity}.

\section{Conversion of near-infrared surface brightness to column density}
\label{sect:method}

Padoan, Juvela \& Pelkonen (\cite{Padoan2006a}) presented a method for
transforming observations of near-infrared surface brightness into estimates
of column density. The main points are repeated below.  We assume that the
cloud is illuminated by an isotropic radiation field, the observed intensity
can be attributed to scattering from dust particles, and observations
of a few near-infrared bands are available. Because scattering traces only the dust column
density, a constant dust to gas ratio is assumed when results are transformed
into total gas column density. The accuracy of these assumptions is discussed
later in this paper.

\begin{figure} 
\resizebox{\hsize}{!}{\includegraphics{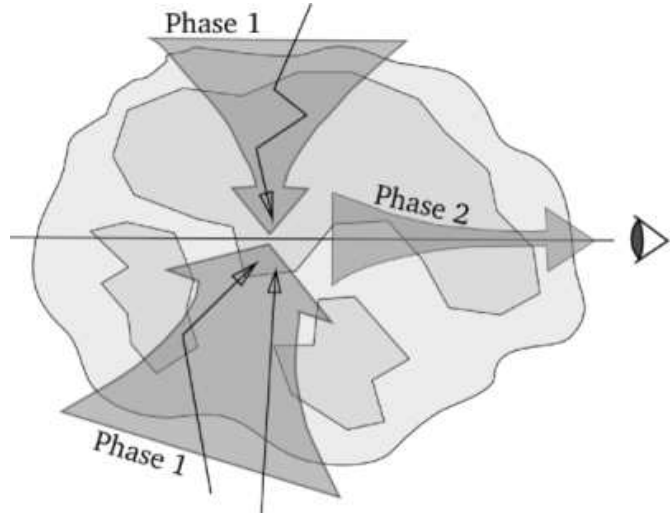}}
\caption{
A schematic view of the creation of the observed surface brightness of
scattered light (see text). The figure shows an inhomogeneous cloud with
arrows indicating the flow of photons. In the first phase external radiation
is transported onto a selected line of sight where it is scattered toward the
observer. Photons reach this line possibly after several scatterings and
preferably through regions of low density. In the second phase radiation 
propagates out from the cloud along the selected sightline. 
} 
\label{fig:scheme}
\end{figure}

In the optically thin case the intensity of the scattered light is directly
proportional to the column density. When the optical depth becomes close to
one, the surface brightness starts to saturate. In the schematic view of
Fig~\ref{fig:scheme} we divide the scattering process into two phases. In the
first phase the external radiation enters the cloud and is transported onto
the selected sightline where it finally scatters toward the observer. In the
second phase the radiation is propagated along the selected sightline toward
the observer. During this phase the intensity is decreased by both absorption
and scattering. This schematic view is useful to understand the reason why our 
method is based on a formalism (e.g. Eq.~\ref{eq:c1}) that corresponds 
to the radiation transfer along an individual line of sight, hence accounts
only for the second phase. The reason is that the second phase is more important 
than the first for the relation between the surface intensity and the column 
density, as the first phase cannot generate strong radiation field variations
within the cloud tightly correlated with the line-of-sight column density.

Because the incoming radiation field is attenuated during the first phase in 
a way that depends on the density structure of the cloud, different locations 
in the cloud could experience different radiation field intensities and this 
could affect the relation between the observed intensity and the dust column 
density, but this effect is relatively small for two reasons. First, a volume 
element is illuminated from all directions. In an inhomogeneous cloud the radiation 
propagates more freely through low density regions so that inside the cloud the
field can be much stronger than what would be expected based on the mean
optical depth. Second, photons can reach a given line of sight also after some
scatterings. Most scatterings are in the forward direction so the radiation
can again better reach deeper cloud regions. Both effects make the field
strength resulting from the first phase relatively uniform within the cloud. 
During the second phase, instead, the attenuation depends on the optical 
depth (column density) of a particular line of sight and every absorption and 
scattering event decreases the observed intensity.

In the near-infrared the albedo of dust grains is close to one half (Lehtinen
\& Mattila \cite{Lehtinen1996}), so the scattering is a significant component
of the total extinction. Taking into account the cloud inhomogeneity, the
effect of the attenuation during the first phase is perhaps half of the effect
during the second phase and, therefore, gives a smaller contribution to the 
relation between surface brightness and line-of-sight column density. The
uncertainty in our method due to this small contribution from the first
phase is studied with radiative transfer simulations.

The relation between surface brightness and column density follows from
the solution of the radiative transfer problem. As explained above, the
observed intensity is determined mainly by radiative transfer effects during
the second phase. This suggests that a radiative transfer equation written for
an individual line of sight should provide a good functional form for the solution.
In the case of a homogeneous medium this results in the equation
\begin{equation}
I_{\nu} = a\, (1 - {\rm e}^{-bN}),
\label{eq:c1}
\end{equation}
which gives the surface brightness of scattered light, $I_{\nu}$, as a
function of column density, $N$. In this formula, $a$ and $b$ are constants
related to the dust properties and to the incoming radiation field. The
constant $b$ is clearly similar to an absorption coefficient. 
This analytical formula was already found to fit the simulation results
of Padoan et al. (\cite{Padoan2006a}). We will examine these correlations for
a larger set of cloud models in Sect.~\ref{sect:correlation}.
In the limit of
low optical depth Eq.~\ref{eq:c1} gives
\begin{equation}
I = a\,b\,N,
\label{eq:c2}
\end{equation}
the intensity is directly proportional to the column density, and the product
$a\times b$ describes the scattering that takes place per unit column density.
If $b$ is assumed to describe dust properties, the constant $a$ would mostly
reflect the intensity of the incoming radiation. However, the interpretation
of these constants is not this straightforward. In the term $exp(-b\,N)$, the
constant $b$ has the role of an extinction coefficient, while in Eq.~\ref{eq:c2} it
would be a scattering coefficient. In the following we treat $a$ and $b$
simply as empirical parameters. We expect that if either the dust properties
or the radiation field were changed, the values of both $a$ and $b$ would
change.  

At large optical depths one must correct for the non-linearity, i.e., the
saturation of the surface brightness. Eq~\ref{eq:c1}, written for each band
separately, defines a parametric curve in the ($I_{\rm J}$, $I_{\rm H}$,
$I_{\rm K}$)-space.
Each observed ($I_{\rm J}$, $I_{\rm H}$, $I_{\rm K}$)-triplet should
correspond to a point on this curve, and the parameter value $N$ could be
calculated if the constants $a$ and $b$ were known. We emphasize that
Eq.~\ref{eq:c1} simply defines an empirical curve representing the
observations. It gives a good description of the relation between surface
brightness and column density, but is not necessarily the optimal function.

When no independent column density estimates are available, surface brightness
observations can only be used to determine the ratios between the NIR bands.
For example, by writing Eq.~\ref{eq:c1} for the H and K bands, $N$ can be
eliminated and the H band can be expressed as a function of the K band,
\begin{equation}
I_{\rm H} = a_{\rm H} \times (1 - (1 - \frac {I_{\rm K}}{a_{\rm K}})^{\frac 
{b_{\rm H}}{b_{\rm K}}}).
\label{eq:c3}
\end{equation}
The constants $a_{\rm J}$, $a_{\rm H}$, and $a_{\rm K}$ and the ratios
$b_{\rm J}/b_{\rm K}$ and $b_{\rm H}/b_{\rm K}$ can be determined by fitting
this curve to the observations. However, determination of the individual
$b$ constants requires additional information.

We have explained above that the $b$ constants depend mainly on the properties
of the dust grains. In the NIR, dust properties vary only relatively little.
If the $b$ constants are similar in different clouds, one can use values already
established from observations of other similar objects. The $a$ constants
depend mainly on the radiation field that illuminates the cloud. The
uncertainty of the intensity is often large, $\sim$50\%. This affects mostly
the absolute scaling of the column density map. The spectral shape of the
incoming radiation can also vary depending, for example, on the presence of nearby OB
associations. However, if the relative values of the scattering cross sections
are known, the spectrum can be estimated using observations of low extinction
sightlines.

The expected values of the constants $a$ and $b$ can be determined by
radiative transfer modelling, while the observed ratios of Eq.~\ref{eq:c3} can
be used to correct the values of $a$ and the ratios between the $b$ constants. If
a correction is necessary, either the radiation field or the NIR dust properties
differ from their initially assumed values. Cloud properties, the average
column density and the inhomogeneity, are also expected to have some effect.
These will be studied in Sect.~\ref{sect:modelling}.

There is still another, more straightforward way to check the validity of
Eq.~\ref{eq:c1}. When NIR scattering is measured, the observations contain
colour excess data for a large number of background stars. The resolution of
the extinction map may be one or two orders of magnitude worse than the
resolution of the surface brightness maps. However, it is sufficient to check
the parameters of the surface brightness method and even to determine the
values of the individual $b$ constants. It is not clear
whether the colour excess method is intrinsically more accurate. However,
since the error sources and error properties of the two methods are very
different, a comparison should be useful to test for systematic effects and 
to provide error estimates for the derived column density maps.

Once the constants are determined, Eq.~\ref{eq:c1} can be used to convert
surface brightness values into column density. The conversion could be done
for each band separately, by inverting Eq.~\ref{eq:c1},
\begin{equation}
N = -\frac{1}{b} log(1-I_{\nu}/a),
\label{eq:c4}
\end{equation}
In principle, observations at one wavelength would be sufficient. However, the
use of several bands is more secure, and it brings additional information that
can be used in the analysis (see, e.g.,
Appendix.~\ref{sect:detailed_modelling}). In this paper we assume that
observations exist of J-, H-, and K-bands. Because of the noise in the
observations and possible model errors the observed intensities do not fall
exactly on the curve given by Eq.~\ref{eq:c1}. For each map position, we find
on this curve a point that minimizes a least square distance from the observed
($I_{\rm J}$, $I_{\rm H}$, $I_{\rm K}$)-triplet. The corresponding column
density is then obtained from Eq.~\ref{eq:c4}.

\section{Modelling of NIR scattering}  \label{sect:modelling}

In this section we describe the numerical modelling used to study 
the relation between the NIR surface brightness and the column
density. The results are based on three-dimensional magnetohydrodynamic
(MHD) turbulence simulations and Monte Carlo radiative transfer calculations.

\subsection{MHD turbulence simulations}

The density structure of our model clouds is based on six MHD turbulence 
simulations that provide a good approximation of the structure of interstellar 
clouds. 

The first three model clouds -- $A$, $B$, and $C$ -- have a resolution of
128$^3$ computational cells. In these models the turbulence is highly supersonic, 
with sonic rms mach numbers {\sc M}$_{\rm s}\sim$10.  The initial density and magnetic
fields are uniform, and an external random force is applied to drive the
turbulence at a roughly constant rms Mach number. In model $C$ self-gravity
is also included.  Model $A$ has approximate equipartition of magnetic and kinetic
energy, the Alfv\'enic Mach number being {\sc M}$_{A}\sim$1. The models $B$
and $C$ have weaker magnetic fields and the turbulence is strongly
super-Alfv\'enic with {\sc M}$_{\rm A}\sim$10. These MHD simulations are the
same used in Juvela, Padoan \& Nordlund (\cite{juvela2001}) for estimation of
molecular line cooling in inhomogeneous clouds, and the reader is referred to
that article for further details.

\begin{figure} 
\resizebox{\hsize}{!}{\includegraphics{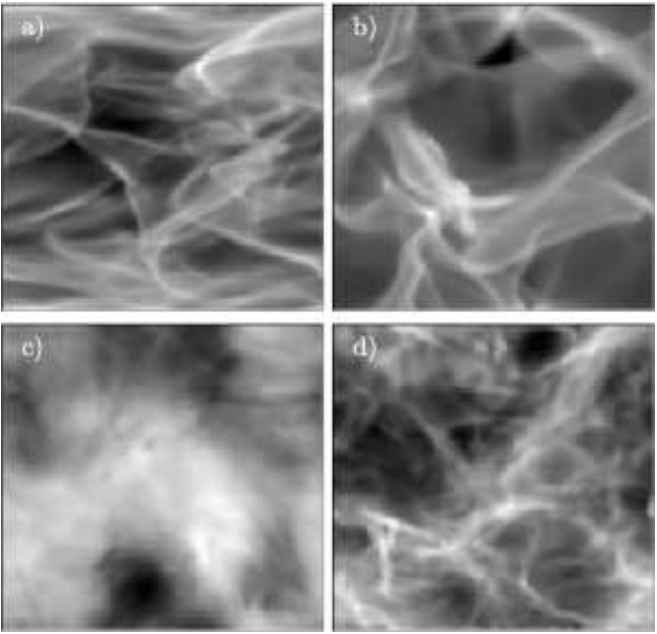}}
\caption{
Examples of the structure seen in the model clouds. The frames $a-d$ show 
column density maps of models $A$, $C$, $D$, and $E$, respectively.  The
colour scales are logarithmic and independent for each frame.
} 
\label{fig:colden_maps}
\end{figure}

The other three model clouds -- $D$, $E$, and $F$ -- are from MHD
simulations on grids of 250$^{3}$ computational cells, but the models
are downsized to 125$^3$ cells for the radiative transfer calculations. The rms
sonic Mach numbers are 0.6, 2.5, and 10.0, for $D$, $E$, and $F$,
respectively.  The initial density and magnetic field are uniform and the gas
is assumed to be isothermal. The initial Alfv\'enic Mach numbers are {\sc
M}$_{\rm A}\sim$10. The volume-averaged magnetic field strength is constant
in time because of the imposed flux conservation. The magnetic energy is
instead amplified. The initial value of the ratio of average magnetic and
dynamic pressures is $\langle P_{\rm m} \rangle _{\rm in} / \langle P_{\rm d}
\rangle _{\rm in}=0.005$, so the run is initially super-Alfv\'{e}nic.  The
value of the same ratio at later times is larger, due to the magnetic energy
amplification, but still significantly lower than unity, $\langle P_{\rm m}
\rangle _{\rm in} / \langle P_{\rm d} \rangle _{\rm in}=0.12$.  The turbulence
is therefore super-Alfv\'{e}nic at all times. Turbulence is again set up as
an initial large-scale random and solenoidal velocity field and is maintained
with an external large-scale random and solenoidal force. Experiments are run
for approximately 10 dynamical times in order to achieve a statistically
relaxed state. The models $D$--$E$ were used, for example, in Juvela,
Padoan \& Jimenez (\cite{cii}) for estimation of [CII] cooling of
inhomogeneous translucent clouds. 
Column density maps from models $A$, $C$, $D$, and $E$ are shown in
Fig.~\ref{fig:colden_maps}.


Supersonic and super-Alfv\'{e}nic turbulence of an isothermal gas generates a
density distribution with a very strong contrast of several orders of
magnitude. It has been shown to provide a good description of the dynamics of
molecular clouds and of their highly fragmented nature (e.g., Padoan et al.
\cite{Padoan+99} \cite{Padoan+01} ,\cite{Padoan+04}). The most important
property affecting the radiative transfer calculations is the degree of
inhomogeneity. In this respect the models provide a wide range of conditions.
Details about the numerical method used in the MHD simulations are
given in Padoan \& Nordlund (\cite{PadoanNordlund1999}). All simulations use
periodic boundary conditions, the effect of which is examined later in
Sect.~\ref{sect:correlation}.

\subsection{Radiative transfer calculations} \label{sect:RT}

The flux of scattered radiation is calculated with a Monte Carlo program
(Juvela \& Padoan \cite{Juvela2003a}, Juvela \cite{Juvela2005a}), where
sampling of scattered radiation is further improved with the `peel-off' method
(Yusef et al. \cite{Yusef94}). During each run, photons are simulated at one
wavelength, and the scattered intensity, including multiple scatterings, is
registered toward a selected direction. The result is an intensity map where
the pixel size corresponds to the cell size of the model cloud. The maps size
is therefore either 125$\times$125 or $128\times 128$ pixels. Maps are
calculated for the three NIR bands J, H, and K (1.25, 1.65, and 2.2\,$\mu$m),
for three directions perpendicular to the faces of the cubic model cloud
(directions $X$, $Y$, and $Z$) and for two diagonal directions ($D1$ and
$D2$). The final maps are averages of several independent runs, allowing
us to control the level of the numerical errors. In the following the
uncertainties are characterized with rms values,
\begin{equation}
  rms(x) = \sqrt{ \sum x_i^2 / N },
\end{equation}
where $x_i$ form a sample of $N$ independent observations of the variable $x$.
In the simulations the Monte Carlo noise, i.e., the rms-value of $\Delta I/I$
resulting from the simulation procedure itself is in all cases below 2\%.
Because of the moderate optical depths and the use of both forced first
scattering and peel-off methods, the errors are relatively uniform, the
relative uncertainty still being somewhat higher in regions of low column
density. In simulations of actual observations, additional noise is added
to these maps in order to simulate the measurement errors. At low intensities
the added noise component is always large compared with the Monte Carlo noise.
At high intensities the Monte Carlo noise becomes dominant, but is still below
$\sim$2\% when the surface brightness exceeds the average intensity of the
map.

In the radiative transfer calculations the model clouds can be scaled
arbitrarily, and the obtained surface brightness is affected only by the 
column density. As an example, we scale all the model clouds
to a size of $L=1.0$\,pc, and rescale the densities to obtain model clouds of
different opacity. In the case of the 125$^3$ cell models $D$--$F$ one cell
would correspond to 0.004\,pc (825\,AU) and, if that cloud were at the distance of
415\,pc, the pixel size would be one arc second.

\begin{figure} 
\resizebox{\hsize}{!}{\includegraphics{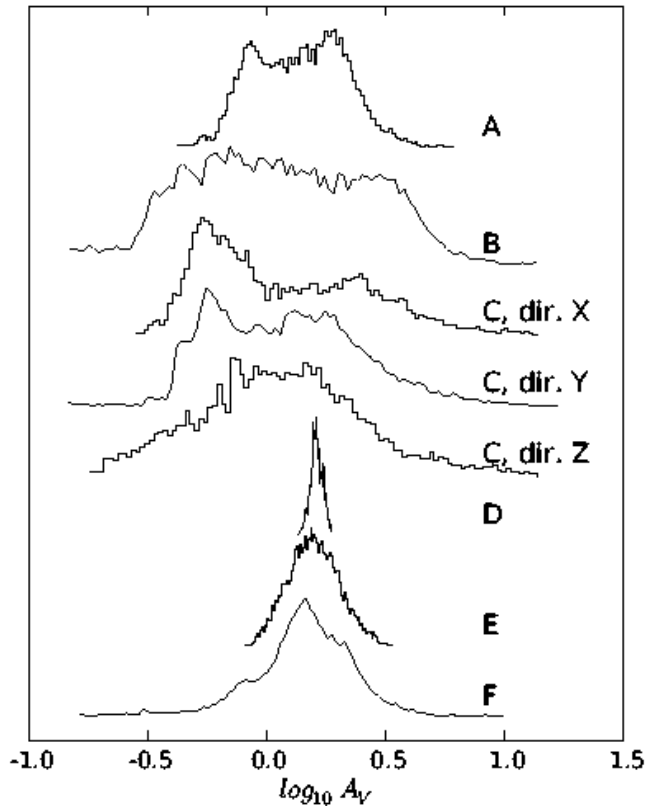}}
\caption{
Histograms of $A_{\rm V}$ values in the six model clouds with average visual
extinction $<A_{\rm V}>=1.6$\,mag. For model $C$ the distributions are plotted
for three orthogonal directions of observations. For the other models
distributions are shown only for one direction. 
} \label{fig:av_distributions}
\end{figure}

Initially we assume a mean density of $10^3$\,cm$^{-3}$. This results in an
average column density of 3.1$\times 10^{21}$\,cm$^{-2}$ which, in the case of
the normal Milky Way dust ($R_{\rm}=3.1$), corresponds to 1.6\,mag of visual
extinction. The actual range of extinctions depends on the inhomogeneity of the
models. Fig~\ref{fig:av_distributions} shows the $A_{\rm V}$ distributions for
the six model clouds in the case of $<n>=10^3$\,cm$^{-3}$. For example, in the
model $D$ practically all lines of sight have extinction below 3 magnitudes,
while in the model $C$ the maximum extinction is over 17$^{\rm m}$. Later we
consider also some models with twice as high density and $<A_{\rm
V}>=3.2$\,mag. In the figures we will use visual extinction instead of column
density. Because the dust properties are constant within each model, the ratio
between $A_{\rm V}$ and true column density is also always constant.

\begin{figure} 
\resizebox{\hsize}{!}{\includegraphics{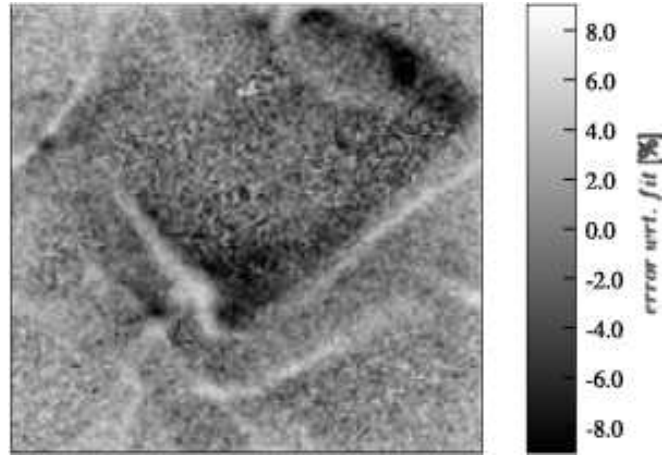}}
\caption{
A map of the difference between the simulated surface brightness and the
prediction based on a curve fitted according to Eq.~\ref{eq:c1}. The cloud is
model $C$ with $<A_{\rm V}>=1.6$. The figure shows the shadowing effects
that are caused by regions of moderate optical depth.
} \label{fig:simu_errors}
\end{figure}

In the simulations the model cloud is illuminated by an isotropic radiation
field, with intensities calculated according to Mathis et al.
(\cite{Mathis83}) model of the ISRF. However,
since the observed surface brightness is directly proportional to the
intensity of the incoming radiation, the simulations can be re-scaled for any
spectrum of the background radiation. Based on DIRBE data
Lehtinen \& Mattila (\cite{Lehtinen1996}) estimated the near-infrared
intensity of the ISRF to be higher by some 50\%. In our analysis this higher
intensity level would only affect the estimated observing times, which would
be shorter by a factor of $\sim$2.

Dust properties are based on Draine (\cite{Draine2003a}) and we use the data
files available on the web \footnote{{\tt
http://www.astro.princeton.edu/$~$draine/dust/}}. In most cases we assume
normal Milky Way dust for which the ratio of total to selective extinction is
$R_{\rm V}=A_{\rm V}/E(B-V)=3.1$. For the scattering function we use
the tabulated scattering phase functions that are available on the web. In
Sect.~\ref{sect:dust} we will examine the effect of dust with $R_{\rm V}=4.0$
or $R_{\rm V}=5.5$, and the effect of spatially varying dust properties. In
those cases the scattering calculations are based on the Henyey-Greenstein
scattering functions (Henyey \& Greenstein \cite{henyey1941}) and the
asymmetry parameters $g$ calculated for the different $R_{\rm V}$ cases.

\section{Results} \label{sect:tests}

In this section we examine how the correlation between cloud column density
and the observed NIR surface brightness depends on factors like the cloud
structure, the average optical depth and the dust properties. The methods of
Sect.~\ref{sect:method} are used to convert surface brightness measurements
into column density estimates. We study the accuracy of the obtained column
density maps and examine how the results could be improved with detailed
radiative transfer modelling or by the use of colour excess data from
background stars.

\subsection{Correlation between scattered light and column density}
\label{sect:correlation}

If NIR scattered light is to be used for a reliable estimation of the column
density, the relation between these quantities should be universal or, when
differences occur, those variations should be taken into account using the
available observations. In the following we study how the results are
affected by the cloud properties.

\begin{figure} 
\resizebox{\hsize}{!}{\includegraphics{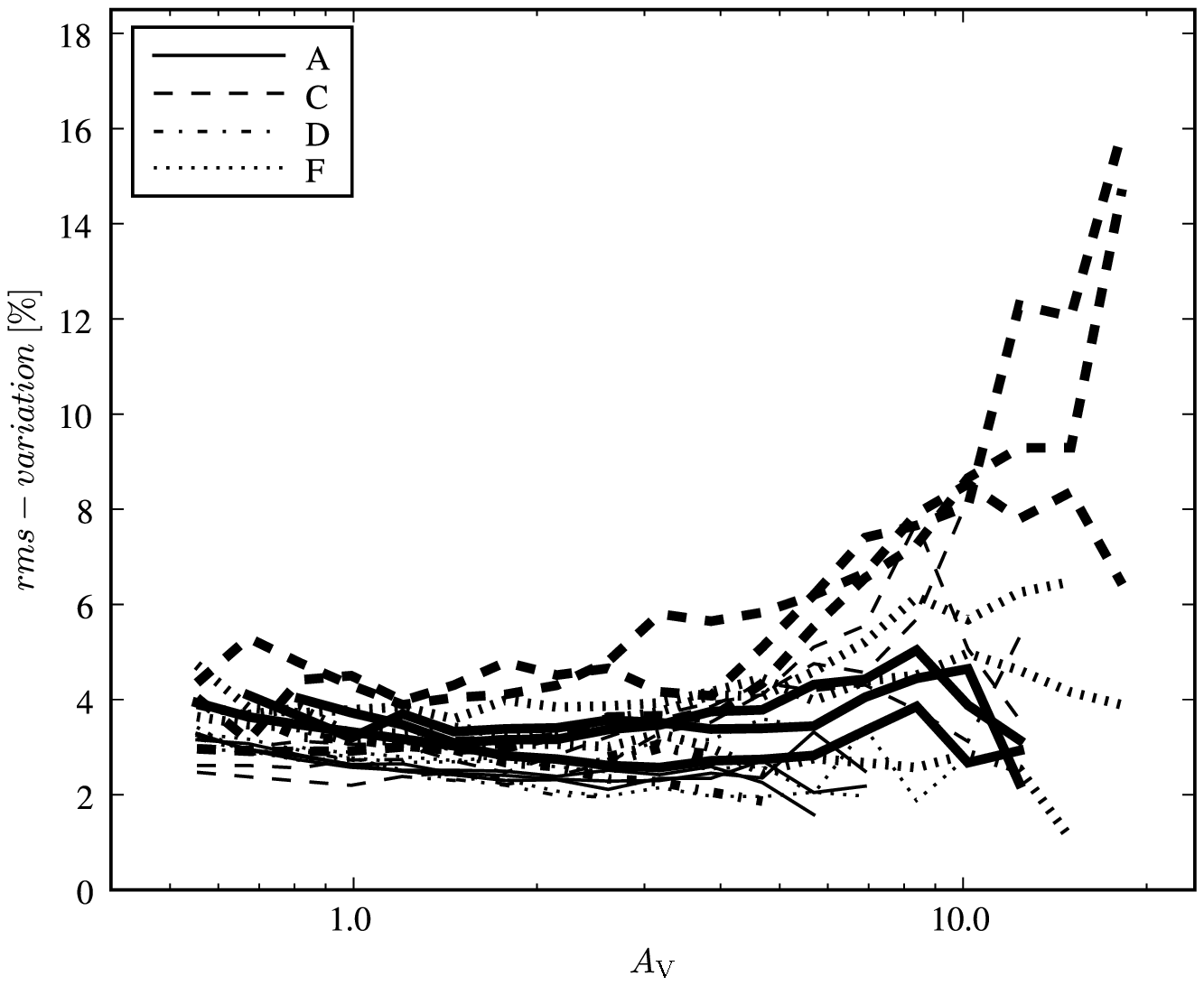}}
\caption{
Relative rms-variation between between the simulated $H$-band surface brightness
and the average curve based on Eq.~\ref{eq:c1}, for models $A$, $C$, $D$, and $E$, 
where the mean visual extinction is scaled to either $A_{\rm}=1.6$ (thin lines) 
or $A_{\rm}=3.2$ (thick lines). For each model separate curves are shown 
corresponding to observations from three different directions ($X$, $Y$, and $D1$). 
This plot illustrates the amount of variation in the surface brightness of scattered 
light for a given line-of-sight column density. 
} 
\label{fig:deviation_vs_Av}
\end{figure}

Figure~\ref{fig:simu_errors} shows the difference between simulated
surface brightness and predictions of Eq.~\ref{eq:c1} that best fits the
points ($N$,$I_{\nu}(H)$). The differences are dominated by systematic
features that are connected with the cloud structure (see
Fig.~\ref{fig:colden_maps}b) and arise from the attenuation of the radiation
field before reaching the line of sight in question. This is most noticeable
in the case of thick filaments. Toward the cloud centre the filaments appear
darker than on the outer side that is subjected to a stronger radiation field.
When the method of Sect.~\ref{sect:method} is applied, this effect will
probably dominate the pixel-to-pixel errors in the derived column density
maps. In Fig.~\ref{fig:deviation_vs_Av} the relative rms-errors,
$rms(\Delta I_{\nu}/I_{\nu})$, are plotted for selected models as functions of
the true $A_{\rm V}$ values.

\begin{figure} 
\resizebox{\hsize}{!}{\includegraphics{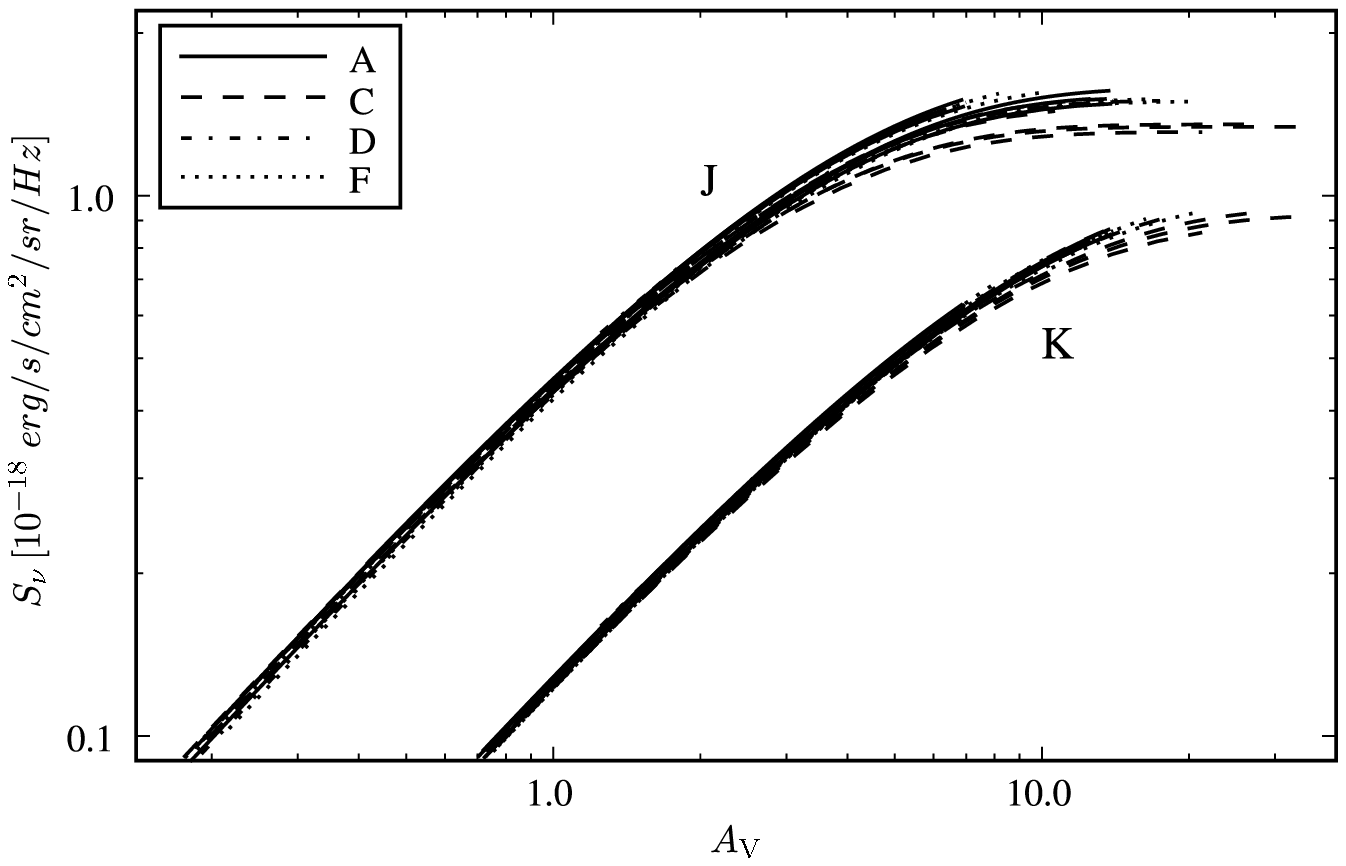}}
\caption{
Curves of Eq.~\ref{eq:c1} fitted to data from selected models (models as in
Fig.~\ref{fig:deviation_vs_Av}). The lower curve is for the $K$ band and the
upper curve for the $J$ band. Each curve is drawn for the range of $A_{\rm V}$
values found in the corresponding model and direction of observations.
} \label{fig:fitted_curves}
\end{figure}

A second source of uncertainty is the fact that parameters of Eq.~\ref{eq:c1}
might vary from source to source in an unpredictable fashion. This could
produce more systematic errors, such as a bias in all column density values or
in the ratio between low and high column densities. We will later study to
what extent such variations can be corrected using the observations
themselves. Here we consider only the amount of variation caused by the
$A_{\rm V}$ and the cloud structure. Figure~\ref{fig:fitted_curves} shows for
some models the fitted curves of Eq.~\ref{eq:c1}. 
The deviation from the average behavior is largest for the most optically
thick model $C$. In this case the maximum extinction is high, over 30
magnitudes. The shadowing produced by such optically very thick regions
is visible even at lower $A_{\rm V}$ as relatively lower surface brightness.

\begin{figure} 
\resizebox{\hsize}{!}{\includegraphics{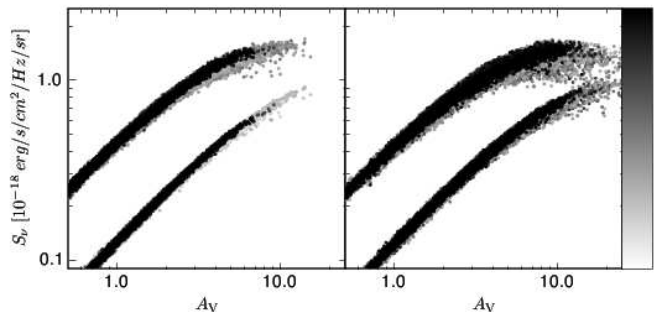}}
\caption{
Scatter plots of surface brightness versus $A_{\rm V}$. The plot
combines data from all six model clouds and three directions of observation
($X$, $Y$, and $D1$). The two frames correspond to average visual
extinction $<A_{\rm V}>=$1.6 and 3.2. The colour scale is linear with respect
to the density of points.
} \label{fig:total_scatter}
\end{figure}

In Fig.~\ref{fig:total_scatter} shows scatter plots of surface brightness 
versus $A_{\rm V}$ for data combined from all six model clouds and three 
viewing directions ($X$, $Y$, and $D1$). The two frames correspond to the 
two different values of average extinction. The figure
shows the total expected dispersion between surface brightness 
and $A_{\rm V}$. It therefore illustrates the accuracy of a column
density estimate based on the surface brightness of NIR scattered light.

In real clouds high densities are limited to the clouds inner parts and are,
therefore, never subjected to full, unattenuated external field. On the other
hand, the MHD turbulence simulations employ periodic boundary conditions and large
densities may be found near the ``surface'' of our model clouds. Furthermore,
when we look at a model perpendicular to its sides (direction $X$, $Y$, or
$Z$), we have sightlines with high densities on both the front and back
side of the cloud. This means that in our simulations the scatter in the relation
between surface brightness and column density is larger than in real clouds. We
examined the magnitude of this effect for all the models with $<A_{\rm
V}>=3.2^{\rm m}$. 
We masked out five-pixel-wide map borders and sightlines where the
density on the front surface exceeded one third of the average density of the
cloud. The scatter with respect to the analytical fit of Eq.~\ref{eq:c1}
decreased by less than $\sim$10\%, showing that edge effects will have only a
small influence on the results.

\subsection{Accuracy of the estimated column density} \label{sect:accuracy}

In this section we use the method of Sect.~\ref{sect:method} to convert
simulated surface brightness maps into estimates of column density. We examine
the effects that column density distribution and dust properties can have on
such estimates. 

There are several possible ways of using Eq.~\ref{eq:c1}. One
could use either a generic set of parameters $a$ and $b$, or try to
improve the parameter values using the observations themselves. For example,
the correct ratio of $a$-parameters can be determined directly from
observations at low column densities. In practice such a procedure might be
necessary, because, for example, the spectrum of the illuminating radiation field
might not be known beforehand with sufficient accuracy. These possibilities
are examined further in Sect.~\ref{sect:extinction} and in
Appendix~\ref{sect:detailed_modelling}.  In this section we use a
single set of values of the $a$ and $b$ constants that are obtained as the average 
of all model runs. As shown by Fig.~\ref{fig:fitted_curves}, there is little
variation between the different models, except for the model $C$ with $<A_{\rm
V}>$=3.2, where the presence of optically thick knots has changed the relation
even for sightlines with $A_{\rm V}\sim$10. This extreme model was excluded from
the calculation of the average parameters. 

We will also consider the effect of random noise to simulate the effect of
observational errors, assuming signal to noise ratios of 20, 15, and 7 for the
J, H, and K bands. We will take the ISAAC/VLT instrument as an example.
Assuming a 0.6$\arcsec$ resolution the total integration time (on-source)
is approximately 50 hours. In the following we assume that one pixel in the simulated
maps corresponds to 0.3$\arcsec$. The S/N ratio {\em per pixel} is lower by a
factor of two, but in the subsequent tests we smooth the maps down to the
resolution of two pixels. Consequently, in the tests the actual noise
corresponds to the S/N ratios quoted above. The ratios were calculated for the
expected average surface brightness of a cloud with $<A_{\rm V}>=1.6$. In the
denser regions and in the more optically thick clouds the relative noise will be
significantly lower. Compared with Padoan et al.~\cite{Padoan2006a} our
integration times are longer, mainly because of the lower average column
density of the models.

The accuracies derived in this section reflect the intrinsic variation between
clouds with different density structure and different optical depth, and the
effect of observational errors.

\subsubsection{Effect of column density distribution} \label{effect:colden}

\begin{figure} 
\resizebox{\hsize}{!}{\includegraphics{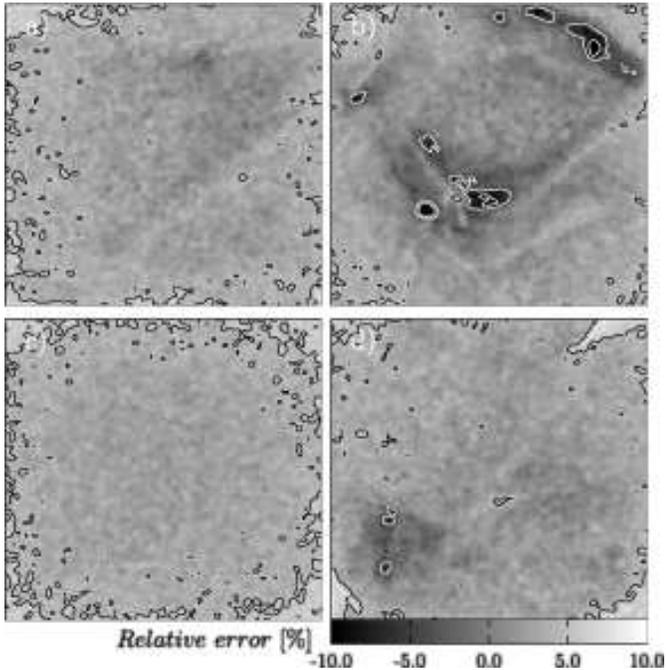}}
\caption{
Maps of relative error of column density estimates in the case of the models
shown in Fig.~\ref{fig:colden_maps}. The average extinction is 1.6 magnitudes
(see Fig.~\ref{fig:av_distributions}), the input maps are convolved with
a beam with fwhm equal to two pixels, and no observational noise is added. The
contours are drawn at the levels of $\pm$5\% and $\pm$10\%. The plots show the
total projected area of model clouds.
} \label{fig:observed_maps_models}
\end{figure}

Fig.~\ref{fig:observed_maps_models} shows some maps of the error in the
column density estimates when no observational noise is added. The input maps
were convolved with a beam with fwhm equal to twice the pixel size so that the
estimated Monte Carlo noise is brought down close to one per cent. Because the S/N
ratio is the same for all bands, the accuracy is limited mostly by the
saturation of the surface brightness and can be improved by giving a larger
weight to the longer wavelength data. In this case the optimal ratio was
1:3:12 for the weighting of J-, H-, and K-band data. In the case of actual
observations, the different S/N ratios will change the optimal ratio. One would
also expect the overall rms error to be reduced if the weight of the shorter
wavelengths were decreased according to the extinction. In practice this was
found to yield only little improvement, and the column density predictions
were calculated using the constant weighting ratio. The errors were largest in
the case of model $C$, where, after the smoothing of the maps, the maximum
visual extinction was 14 magnitudes. The maximum error was 23\%. The average
relative rms error was only 1.7\%, large part of this number still being noise
from the simulation procedure. This is to be expected, since the majority of
sightlines are optically thin and the surface brightness is directly proportional
to the column density. In model $C$ the largest deviations are found in
dense filaments. The other clouds are more homogeneous, and the errors reflect
large scale gradients in the strength of the radiation field: The column density
is systematically over-estimated close to the cloud borders. 

\begin{figure} 
\resizebox{\hsize}{!}{\includegraphics{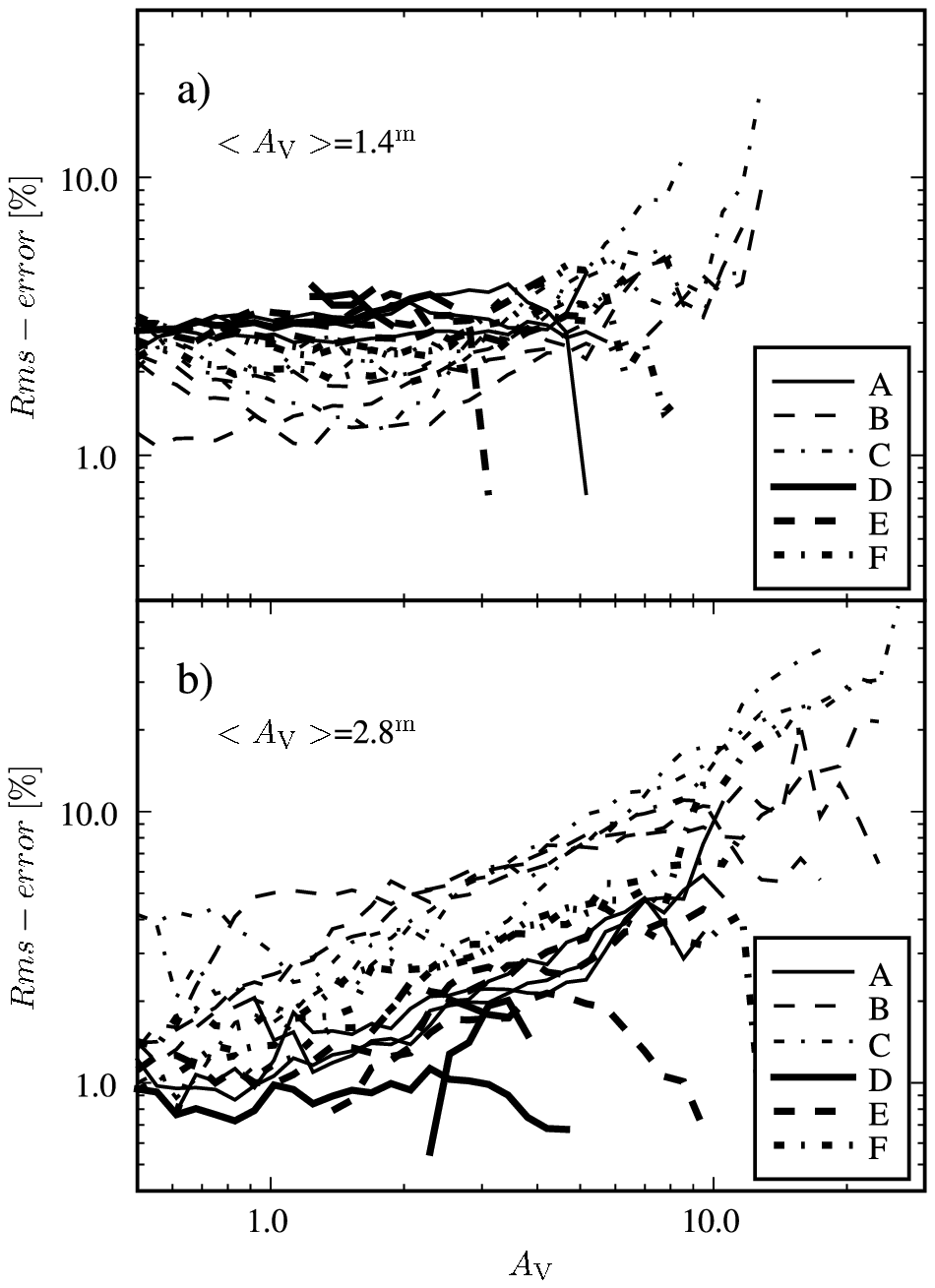}}
\caption{
Rms-error of column density estimates as a function of the line-of-sight
visual extinction. The plot includes results for all six model clouds, for
three different directions of observation ($X$, $Y$, and $D1$). The frames
correspond to the two values of average visual extinction.
} \label{fig:error_vs_Av}
\end{figure}

In Fig.~\ref{fig:error_vs_Av} we show the average rms-error of the column
density estimates as a function of $A_{\rm V}$. As before, the predictions
are calculated using the average values of the parameters $a$ and $b$. In
models with $<A_{\rm V}>$=1.6 the errors remain mostly below 10\% with the
exception of model $C$ where, as already mentioned, the largest error is
$\sim$23\%. In the optically thicker clouds, $<A_{\rm V}>=3.2^{\rm m}$ , the
errors are larger for all sightlines, as the shadowing by optically thick
regions causes fluctuations in the strength of the radiation field. However,
below $A_{\rm V}\sim 10^{\rm m}$ the errors are at most 10\%. Above $A_{\rm
V}\sim 20^{\rm m}$ the average errors increase close to 50\%.

\begin{figure} 
\resizebox{\hsize}{!}{\includegraphics{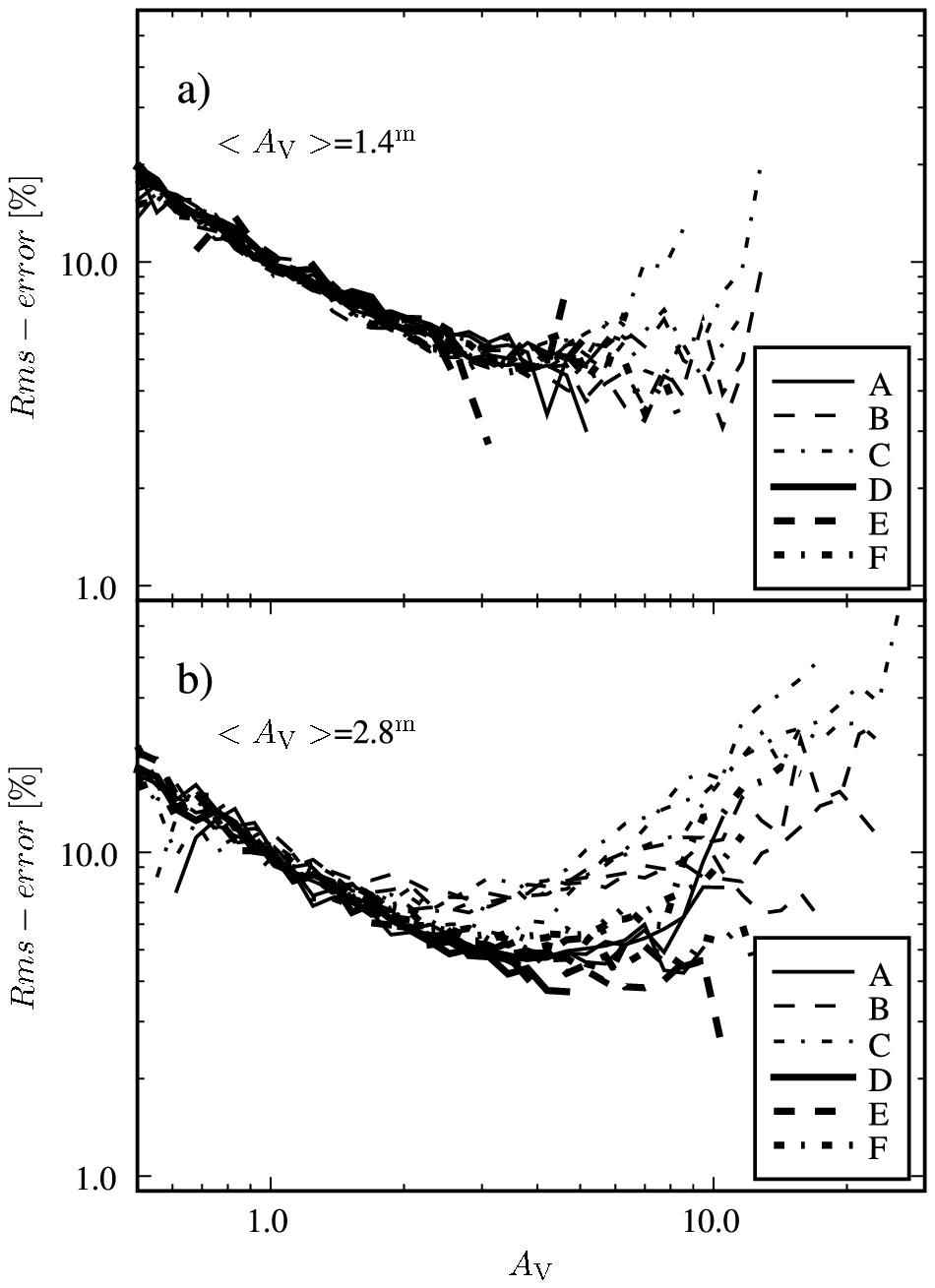}}
\caption{
Rms-error of column density estimates as a function of $A_{\rm}$. 
The plot includes results for all six model clouds and
for three different directions of observation ($X$, $Y$, and $D1$). 
Observational noise was added to the input maps (see text). The 
frames correspond to the two values of average visual extinction.
} 
\label{fig:noisy_error_vs_Av}
\end{figure}

\begin{figure} 
\resizebox{\hsize}{!}{\includegraphics{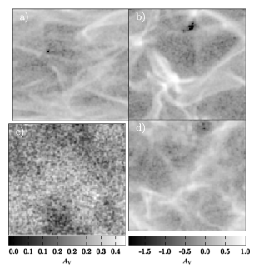}}
\caption{
Estimated column density maps of the four models ($A$, $C$, $D$, and $E$) of
Fig.~\ref{fig:colden_maps}. The column densities are calculated based
on surface brightness maps where observational noise is included (see text).
The images show the logarithm of the column density, transformed into $A_{\rm
V}$. The frames $a$, $b$, and $d$ share a common colour scale. The model $D$
has a much smaller range of column densities (see
Fig.~\ref{fig:av_distributions}), and the frame $c$ has a different colour
scale.
} 
\label{fig:noisy_maps}
\end{figure}

Next, we add noise to the surface brightness maps to simulate the observational
uncertainties. The S/N ratios are 20, 15, and 7 for the J, H, and K bands
respectively, when calculated for the average surface brightness of the
$<A_{\rm V}>$=1.6 maps and the maps are smoothed to a resolution of two
pixels. The weighting of the different bands is kept the same as before. The
results are shown in Fig.~\ref{fig:noisy_error_vs_Av}. The main effect of the
added noise is a decrease in the accuracy at small column densities. The
relative errors are $\sim$10\% at $A_{\rm V}=0.5^{\rm m}$, decrease until
$A_{\rm}\sim 4^{\rm m}$, and again increase at higher column densities as in
Fig.~\ref{fig:error_vs_Av}. A good signal-to-noise ratio should be
particularly important at high opacities where, due to the saturation of the
surface brightness values, a small change in surface brightness corresponds to
a large change in column density. However, in these regions the S/N ratios are
already very high and, beyond $A_{\rm V}\sim 8^{\rm m}$, 
Fig.~\ref{fig:error_vs_Av} and Fig.~\ref{fig:noisy_error_vs_Av} are
practically identical. Figure~\ref{fig:noisy_maps} shows maps of the
estimated column density maps for the four models of
Fig.~\ref{fig:colden_maps}.

\begin{figure} 
\resizebox{\hsize}{!}{\includegraphics{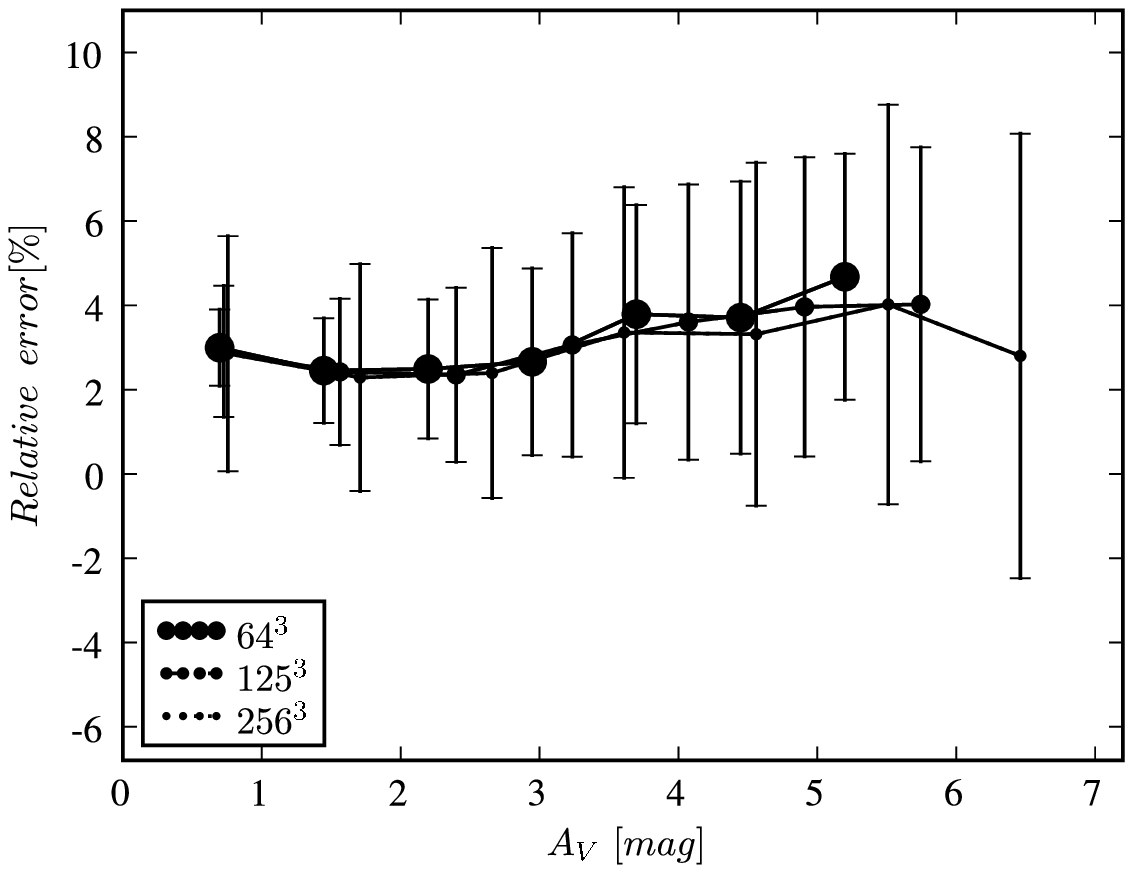}}
\caption{
The errors of the estimated column densities when model $F$ is
discretized into 64$^3$, $125^3$, or $256^3$ cells. The estimates are based on
average parameters derived from all models with $<A_{\rm V}>=1.6^{\rm m}$ and
$<A_{\rm V}>=3.2^{\rm m}$. The points show the average bias in different
$A_{\rm V}$ intervals and the errorbars correspond to the 1-$\sigma$ variation
within each interval. At low $A_{\rm V}$ part of the differences is caused by
the Monte Carlo noise of the simulations, which was higher for models of
higher resolution.
} 
\label{fig:reso}
\end{figure}

The column density distribution also depends on the resolution of the
observations or, in our case, the resolution of the simulations. The
discretization of the cloud models could affect the results if individual
cells were optically thick {\em and} if a finer discretization resolved
significant density variations within the original cells. The largest density
contrast is found in model $C$. When the cloud has an average visual
extinction of $<A_{\rm V}>=3.2^{\rm m}$, the $J$-band optical depth of the
densest cell is $\sim$1.3 and 99.99\% of the cells have optical depth below
$\tau_{\rm J}$=0.5. The maximum optical depth is smaller for the other bands
and models and, naturally, for the clouds scaled to $<A_{\rm V}>=1.6$.
Therefore, discretization is not expected to produce errors that could affect
our results. This was still checked using the model $F$. The original MHD
simulation was done on a grid of 256$^3$ cells and based on that we
constructed three models consisting of 256$^3$, 125$^3$, and 64$^3$ cells.
Radiative transfer simulations were repeated for all the three model clouds,
and Fig.~\ref{fig:reso} shows the errors when column densities are estimated
at the full resolution of each model.  The larger differences at low $A_{\rm
V}$ are caused by a difference in the Monte Carlo noise of the simulations.
Otherwise, the accuracy of the column density determination is not seriously
affected by the numerical resolution. 

\subsubsection{Effect of radiation field and dust properties} \label{sect:dust}

In deriving the column densities we have so far assumed that model parameters
$a$ and $b$ have already been determined from other models illuminated with
similar external radiation field and containing dust with exactly the same
scattering properties. In reality, both radiation field and dust properties 
can be subject to significant variations. 

If the total level of the incoming radiation is underestimated, the column
density is correspondingly overestimated. Below $A_{\rm V}\sim 10^{\rm m}$
the surface brightness is directly proportional to the external field so that
an error in the assumed intensity of the radiation field results in similar
relative error in the column densities. This is a constant factor and does not
affect relative accuracy between map positions. An error in the assumed
spectrum, i.e., in the intensity ratio between different bands, can result in
different kinds of errors, especially if the bands are weighted differently,
according to the local S/N-ratios. However, observations of optically thin
sightlines can be used to correct for changes in the spectrum and, partially,
changes in dust properties.

We repeated simulations of model $C$ using dust with $R_{\rm V}=4.0$ and
$R_{\rm V}=5.5$. The dust parameters were taken from Draine et al.
(\cite{Draine2003a}). The J- and H-band intensities were plotted against the
K-band intensity, and Eq.~\ref{eq:c3} was fitted to the data. The parameters 
$a_{\rm K}$ and $b_{\rm K}$ were kept constant using the previous values
that were derived assuming $R_{\rm V}=3.1$. In the NIR the shape of the
extinction curve remains rather constant, and the main effect results from
differences in the absolute values of the absorption and scattering cross
sections.
In the optically
thin regime the product $a\times b$ is the slope of the relation between
surface brightness and column density. Since these ratios of $a\times b$ in
different bands can be determined
from the data itself, no a priori assumption needs to be made about the
spectrum of the external radiation, and some of the uncertainty caused by
unknown dust scattering properties is eliminated. However, because the value
of the product $a_{\rm K} \times b_{\rm K}$ must be assumed, the column
density estimates may be at error by a constant factor as discussed above. 
The situation is worse at high column densities because we cannot directly
determine the column density at which the saturation of the surface brightness
values starts. While the ratios between the $b$ parameters can be recovered from
the data, the value of an individual $b$ parameter must be assumed a priori. If the 
dust extinction cross sections
are different from the assumed one, the saturation begins at a
higher or lower column density than expected, and all column density estimates
of high extinction sightlines are correspondingly erroneous. Moreover, if the
radiation field is stronger than expected, the column densities will be
overestimated in the non-linear part of Eq.~\ref{eq:c1}. If the surface
brightness exceeds the expected asymptotic value, represented by the
parameters $a$, no solution can be found. However, in such a case the peak
surface brightness can be used to correct the value of $a_{\rm K}$.

\begin{figure} 
\resizebox{\hsize}{!}{\includegraphics{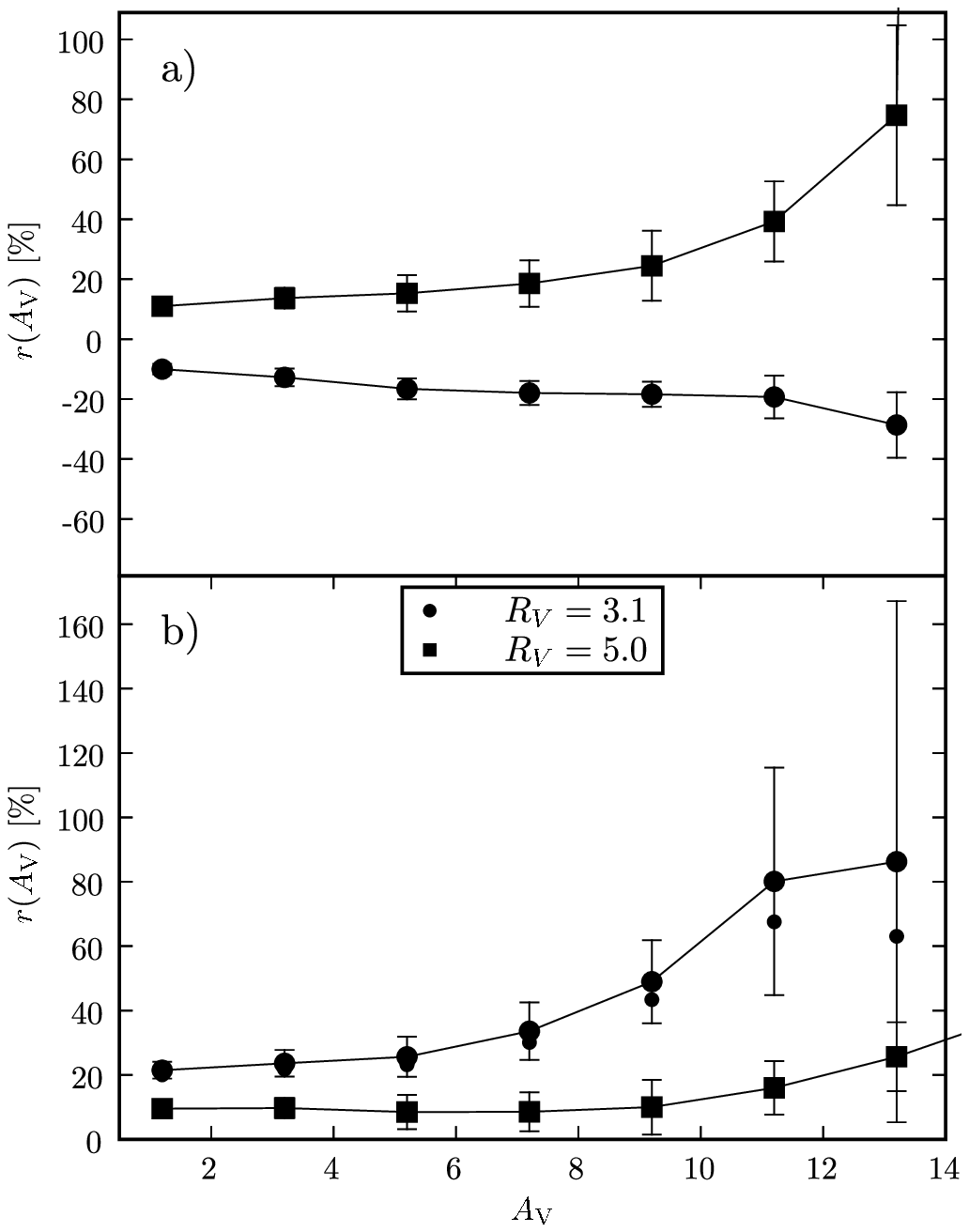}}
\caption{
Effect of a change in dust properties and/or field strength on the
column density estimates. The plots show the bias and scatter in the
estimates when dust properties are assumed to correspond to $R_{\rm V}=4.0$,
while they in reality correspond to either $R_{\rm V}=3.1$ or $R_{\rm V}=5.5$.
In frame $b$ the radiation field is also 30\% higher than assumed in the
column density derivation, and the parameter $a_K$ is re-estimated based on
the maximum surface brightness. The small circles show the result in the case
of $R_{\rm V}=3.1$ without re-evaluation of parameter $a_K$.
} 
\label{fig:dust_vs_av}
\end{figure}

Figure~\ref{fig:dust_vs_av} shows the bias and statistical noise in the cases
of $R_{\rm V}=3.1$ or $R_{\rm V}=5.5$. The $b_{\rm K}$ and
$a_{\rm K}$ parameters correspond to a case $R_{\rm V}=4.0$, and the other
parameters have been re-estimated from the plots of the J- and H-band intensity
versus the K-band intensity. In frame $a$ the radiation field is the same as before.
No additional observational noise was included. The surface brightness of the
scattered light increases with the $R_{\rm V}$ value of the dust. Therefore, for 
$R_{\rm V}=5.5$ the column densities are overestimated and on the non-linear part the
systematic error increases rapidly as the line-of-sight extinction exceeds
$A_{\rm V}\sim 10^{\rm m}$. The maximum observed surface brightness was not
used to correct the value of the parameter $a_{\rm K}$. In the case $R_{\rm
V}=3.1$ the column densities are underestimated, but errors are smaller, being
$\sim$20\% or less for $A_{\rm V}\sim 10^{\rm m}$.

In frame $b$ the radiation field was increased by 30\% relative to the value
assumed in the column density derivation. At higher column densities, and
especially in the case with $R_{\rm V}=5.5$, the column densities would be
either seriously overestimated or no value could be obtained from
Eq.~\ref{eq:c1}. Therefore, the analysis was done using a corrected $a_{\rm
K}$ value that was set to the maximum observed K-band intensity. Such a
correction assumes that maps contain at least one optically thick region.
Because we have assumed that the radiation field is underestimated, the column
densities are overestimated for all sightlines and for both dust models.
However, in the case of dust with $R_{\rm V}=5.5$ the errors are, in fact,
much smaller than before. Apparently the correction of the parameter $a_{\rm
K}$ compensated also for the higher than expected dust scattering efficiency.
On the other hand, for dust with $R_{\rm V}=3.1$ the corrected $a_{\rm K}$
value does not improve the estimates. Without clearly saturated regions the
maximum surface brightness is a poor measure of $a_{\rm K}$. In that case the
parameter values can be corrected only by other methods, for example by
comparison with an extinction map (see Sect.~\ref{sect:extinction}). In
some cases the accuracy can be improved by more detailed modelling of the
observed source. This possibility is discussed further in
Appendix~\ref{sect:detailed_modelling}.

\subsection{Use of extinction measurements} \label{sect:extinction}

Observations of NIR surface brightness will automatically give
photometry for a large number of background stars. The colour excesses of the
stars can be converted into estimates of the clouds column density (Alves,
Lada \& Lada \cite{alves2001}; Lombardi \& Alves \cite{lombardi2001};
Cambr\'esy et al. \cite{cambresy2002}). Each star can be treated as a probe 
of the extinction along individual, very narrow sightlines. 
Usually the results are spatially smoothed to give a map of
extinction at lower resolution. Averaging is required because of the scatter in
the intrinsic colours of the stars. The resolution of the resulting column
density map is much lower than the resolution of the surface brightness maps.
This information can, however, be used to assist the use of surface brightness
measurements. The combination of the two methods is interesting because, while
both rely on absorption and scattering by dust particles, their actual
dependence on the dust properties is different. Furthermore, the colour excess
method is independent of the local radiation field. This makes it possible to
take into account any radiation anisotropies in the studied field. Conversely,
differences in the two column density estimates serve as a sensitive
indicator of variations in dust properties and the radiation field.

In the following we use simplified simulations of the colour excess method
where extinction maps are calculated with the NICER method (Lombardi \& Alves
\cite{lombardi2001}). Stars are placed at random locations behind the model
cloud and their extinction is proportional to the true column density. The magnitude
distribution follows a linear ($mag$, $log N$) relation with a slope of 0.35.
A random error of 0.12 magnitudes is added to each band in order to simulate
photometric errors and the variation in the intrinsic colours of the stars.
The resulting dispersion in the colours is similar to or slightly smaller than
the scatter found in 2MASS data. For simplicity, the same noise was applied to
all stars. Faint stars are removed when their magnitudes exceed some limit
$m_K$ in the K-band, $m_K+0.85$ magnitudes in the H-band, or $m_k+1.5$
magnitudes in the J-band. The $m_K$ limit determines the remaining number of
stars.

\begin{figure} 
\resizebox{\hsize}{!}{\includegraphics{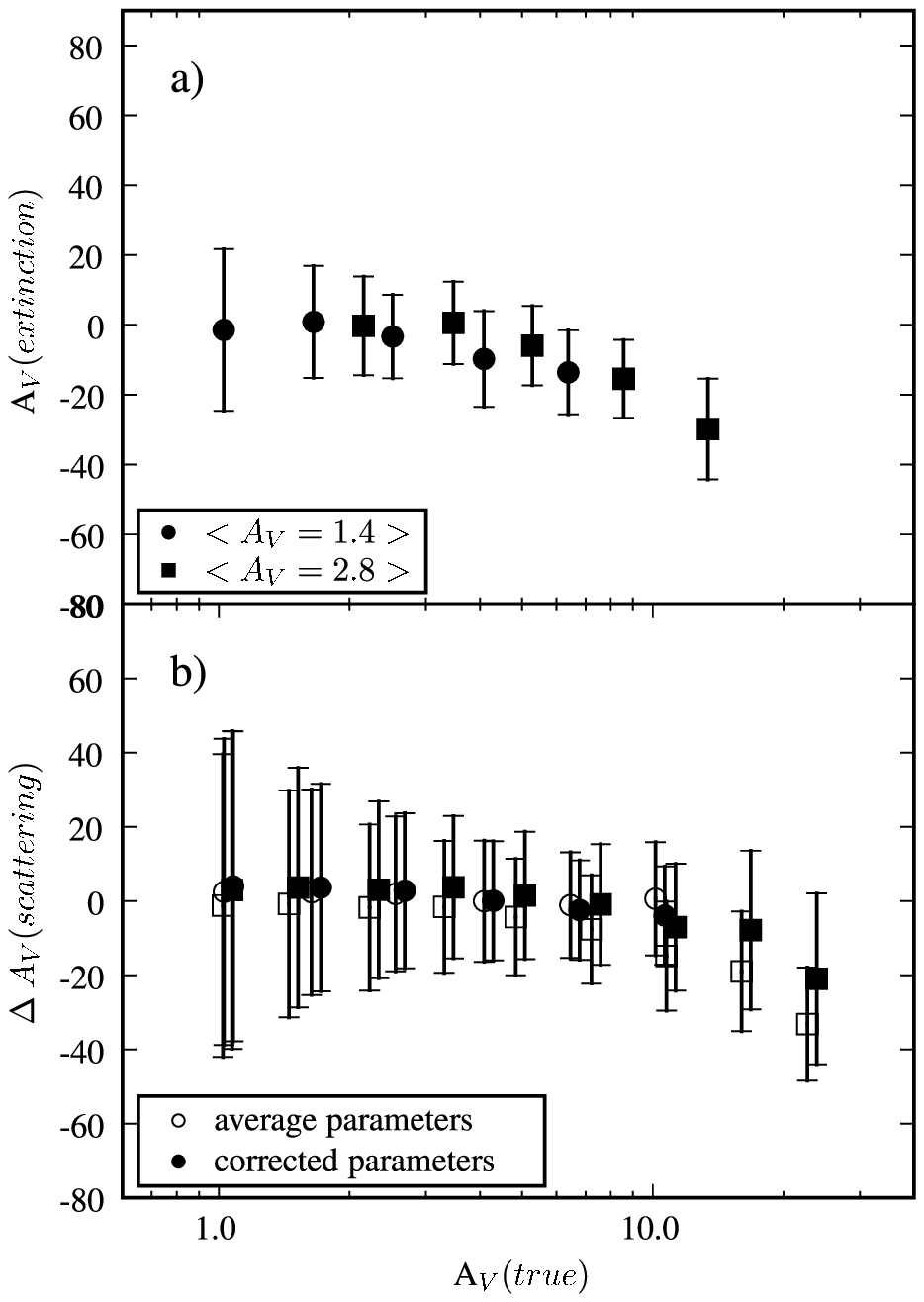}} 
\caption{
{\em Upper frame:} Errors in the simulated extinction map. Plots are
shown for two versions of model $C$, with average visual extinction 1.6
(circles) or 3.2 magnitudes (squares). The resolution of the extinction map
corresponds to a FWHM equal to seven map pixels.
{\em Lower frame:} Accuracy of the column density estimates based on NIR
scattering. Results are shown for the two versions of the model $C$. The
parameters are either the average values from all models (six clouds, three
viewing directions and two scalings of average column density; open symbols)
or were determined by correlating surface brightness against column
densities estimated using the colour excess method (solid symbols).
} 
\label{fig:extinction_para} 
\end{figure}

The comparison of extinction and surface brightness gives a way to determine
all the parameters of Eq.~\ref{eq:c1} directly from observations. This was
tested on model $C$, using the 128$\times$128 maps and adding observational
noise as in the previous section. The upper frame of
Fig.~\ref{fig:extinction_para} shows the accuracy of the extinction maps that
were derived from simulated colour excess observations and were averaged with
a Gaussian with FWHM equal to 7 pixels.  The number of background stars over the
whole map was $\sim$2000.
The lower frame shows the results for the surface brightness method
together with errors calculated for individual pixels.  The parameters $a$ and
$b$ were either average values of all model clouds, including both models with
$<A_{\rm V}>=1.6^{\rm m}$ and $<A_{\rm V}>=3.2^{\rm m}$ (see
Sect.~\ref{sect:RT}), or were estimated through a comparison of surface
brightness and extinction maps. 
The colour excess data did not significantly improve the accuracy of the
surface brightness method. This is not surprising because the average
parameters were based on simulations with correct dust properties and correct
value of the background radiation field. On the other hand,
Figure~\ref{fig:extinction_para} does show that by using the colour excess
information similar accuracy can be reached even without any such a priori
knowledge of the dust properties and of the radiation field. Extinction
measurements are therefore useful to avoid assumptions about dust properties
and radiation field intensity in the application of our method. The column density
map obtained with our method still has a much better spatial resolution that the 
extinction map itself, seven times better in this case.

\begin{figure*} 
\resizebox{\hsize}{!}{\includegraphics{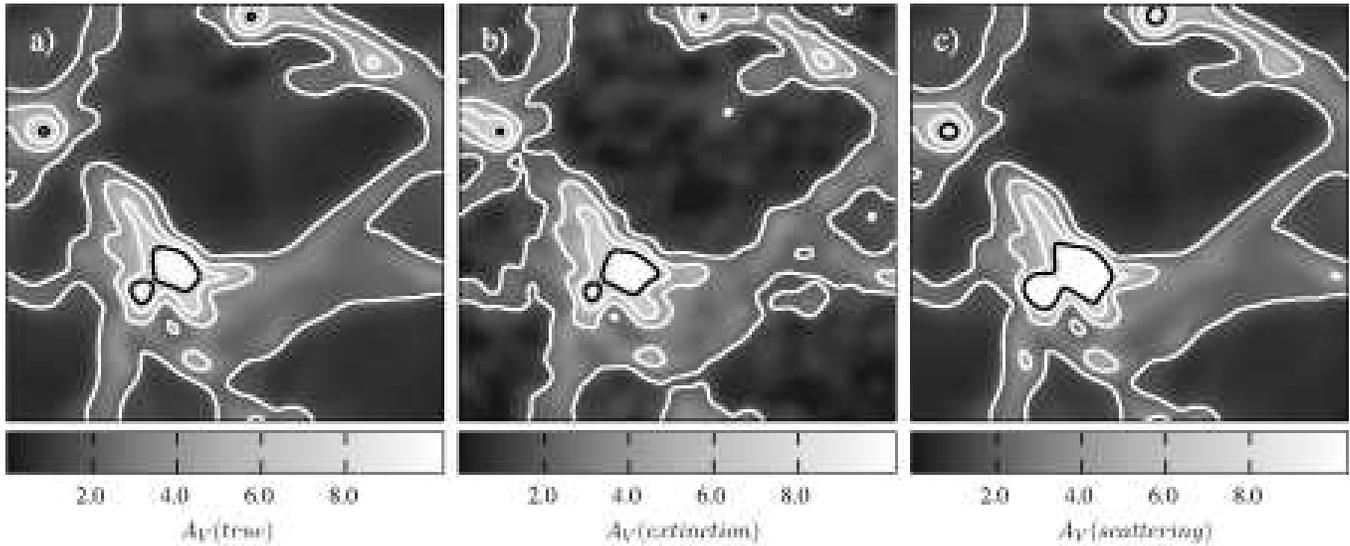}} 
\caption{
Model containing variable dust properties,  $R_{\rm V}=3.1$ in low density
regions and $R_{\rm V}=5.5$ in high density regions. The frames are: {\em a.}
True $A_{\rm V}$, {\em b.} $A_{\rm V}$ based on colour excesses of $\sim$2000
background stars, and {\em c.} $A_{\rm V}$ based on scattered surface brightness.
All maps have been smoothed to a resolution of six map pixels.
} 
\label{fig:bicomponent}
\end{figure*}

In the previous discussion, the colour excess information was used globally,
through the model parameters $a$ and $b$. Alternatively, one could combine the
extinction map and the column density map obtained from our method, at the lower
resolution of the extinction map. Spatial variations in the ratio of the maps could
serve as an indicator of changes in dust properties or in the strength of the
local radiation field.  To test these possibilities we modified the model $C$
($<A_{\rm V}>=1.6^{\rm m}$), first re-examining the effect of changing dust
properties. The fractional abundance of dust with $R_{\rm V}=3.1$ was set to
depend on local density according to the function $f=n/(2000+n)$. Dust with
$R_{\rm V}=5.5$ was given relative abundance $1-f$. Column densities were
estimated with both methods (Fig.~\ref{fig:bicomponent}). The colour excess
method gives very accurately the correct average visual extinction but, with
2000 simulated background stars, the map has significant noise. When the
surface brightness method is applied using parameters tuned for $R_{\rm
V}=3.1$ dust, the column densities of the same regions are significantly
over-estimated. The ratio of the two estimates reflects, therefore, both
resolution effects and the changes in dust properties. The intensity ratios of
the NIR bands would, of course, also directly show that the assumption of
either the dust properties or the radiation field is not correct. As above,
the extinction map is used to re-evaluate the average global parameters for
the surface brightness method. The resulting map is shown in
Fig.~\ref{fig:bicomponent}c.  The highest column densities are slightly
overestimated but the map is morphologically extremely accurate. As before,
the surface brightness data contained observational noise.  In 
Fig.~\ref{fig:bicomponent} all the three maps are smoothed to a resolution of
six pixels, but the signal-to-noise ratios would allow a much higher spatial
resolution for the map that was based on the surface brightness data. 
Extinction measurements could also be used in the creation of a detailed
radiative transfer model of the source (see
Appendix~\ref{sect:detailed_modelling}). That would be particularly useful in
the case of a strongly anisotropic radiation field.

\begin{figure*} 
\resizebox{\hsize}{!}{\includegraphics{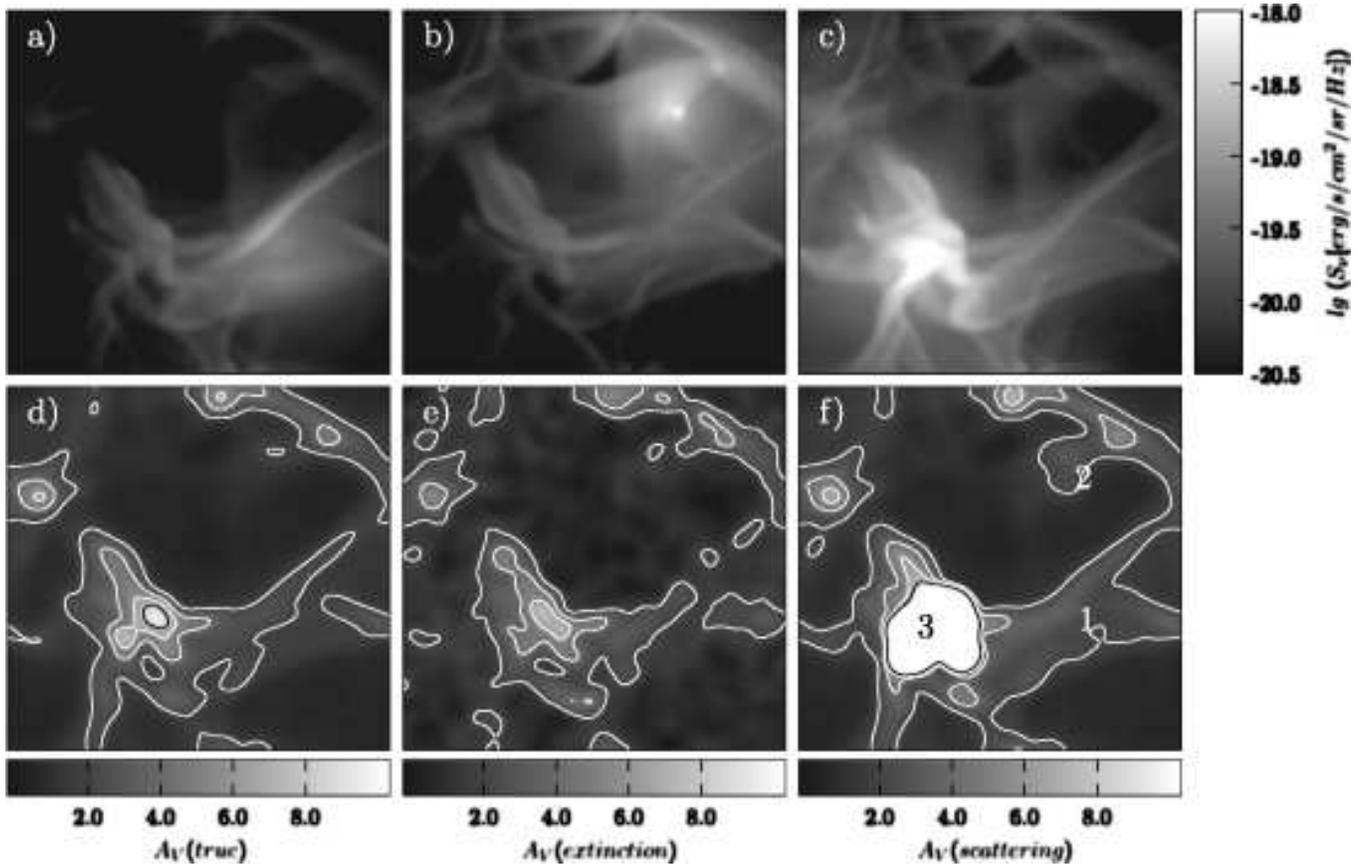}} 
\caption{
{\em Frames a-c:} H-band surface brightness maps for each of the three 
point sources that were added to model cloud $C$.
{\rm Frame d-f:} Column density estimates for model $C$ containing three point
sources. The maps show the true column density, estimates based on NICER
method and estimates based on scattered surface brightness. Note that the
colour scale is logarithmic in the upper frames and linear in the lower
frames.
} 
\label{fig:sources} 
\end{figure*}

Local radiation sources may complicate the derivation of the column densities.
Depending on the spectrum of these sources, it may be difficult to deduce
their presence based on surface brightness data only. This is true especially
if the true column density is low and the surface brightness itself is not
anomalously high. In that case the sources could lead to significant
overestimation of the column density. In the following we examine how such
errors manifest themselves in the estimated column density maps and to what
extent the colour excess data can be used to correct the results. We added to
cloud $C$ three point sources, each with a 4000\,K black body spectrum. The
K-band luminosities were 1.5$\times 10^{19}$, 3.1$\times 10^{19}$, and
6.2$\times 10^{19}$ erg/s/Hz. Figure~\ref{fig:sources}a-c shows the surface
brightness maps for each source separately, excluding direct radiation from
the source. The lower frames show again the true column densities, the NICER
estimates (using 2000 background stars), and the estimates based on scattered
light, all convolved with a beam with FWHM$\sim$6 pixels. In observations, the
presence of the strongest source would be obvious, based on the presence of
the point source, high surface brightness values in excess of the ISRF
intensity, and comparison with the NICER map. The second source can still be
seen by plotting the difference of the two column density estimates. In the
surface brightness method (using default parameters) the estimated column
density at the position of source 2 is twice the correct value. However, the
effect of this source is very local. At the distance of 6 pixels the error is
still $\sim$100\% but falls below 50\% at the distance of 12 pixels. The
weakest source is outside dense regions and therefore does not produce A clear
peak in the surface brightness or estimated column density. There is, however,
a diffuse area ($\sim$1\% of the whole map) where the column density is
overestimated by $\sim$30\%. Because the area has low extinction, $A_{\rm
V}<4$, the noise is relatively high in the NICER map, and it would be
difficult to detect the effect of this source in actual observations.

\section{Discussion}  \label{sect:discussion}

Observations of scattered NIR light provide a means to map large areas of
interstellar clouds with very high spatial resolution. Our simulations showed
that one can reach average pixel-to-pixel accuracy better than 10\% as long as
the maximum extinction is below $A_{\rm V}\sim 15^{\rm m}$. High absolute
accuracy requires that the parameters of Eq.~\ref{eq:c1} can be determined
with similar precision. This requires knowledge of both the radiation field
and the dust properties. Fortunately, the NIR observations will automatically
give colour excess data for a large number of background stars. Comparison of
surface brightness data and the colour excess data provides a direct way to
determine the parameters of Eq.~\ref{eq:c1}. Therefore, the column density
estimation can be carried out with little a priori information. This is
particularly important regarding the radiation field because its intensity is
usually unknown and cannot be easily determined. On the other hand, the NIR
dust properties are generally believed to be rather constant among sources
of similar type. If the conversion between surface brightness and column
density is done with incorrect parameters, the absolute scaling and the
relative values, as a function of position or column density, will be
incorrect. However, the derived column density maps are still usually
morphologically accurate (see, e.g., Fig.~\ref{fig:bicomponent}). 

The use of Eq.~\ref{eq:c1} contains the assumption that the intensity of the
radiation field and the dust properties are constant within the cloud. Spatial
variations of dust properties were examined in Sect.~\ref{sect:extinction}. In
particular, Fig.~\ref{fig:dust_vs_av} showed that although variation between
dust with $R_{\rm V}=3.1$ and $R_{\rm V}=5.5$ can introduce visible changes in
the estimated column density maps, the variations are still only at the 10\%
level. Our simulations showed that as long as the clouds are not very
optically thick, $A_{\rm V}\sim 15^{\rm m}$ or below, the errors caused by
large scale radiation field gradients are $\sim$10\% or below. The effect was
visible mainly towards the edges of relatively homogeneous clouds (see
Fig.~\ref{fig:observed_maps_models}). 

At large scales the spatial variations in the dust properties and the
radiation field can be recognized with the help of background stars. An
extinction map that is based on background stars will have a much lower
resolution than the surface brightness maps. Furthermore, because the
parameters of Eq.~\ref{eq:c1} can only be determined using correlations over
large areas, rapid spatial variations in dust properties may not be taken
fully into account. Nevertheless, the comparison of extinction and surface
brightness maps will give indication of those variations. Some effects can be
recognized and corrected based on surface brightness data itself. This is
discussed further in Appendix~\ref{sect:detailed_modelling}, where we examine
the possibility of improving the accuracy of the column density estimates
through detailed radiative transfer modelling. The effects resulting from
radiation field and dust property variations are examined further in
Appendix~\ref{sect:validity}. 

We have assumed that the observed surface brightness is caused entirely by
dust scattering. The contribution from emission lines is small but the role of
NIR dust emission remains uncertain. At low $A_{\rm V}$ sightlines a
significant part of the surface brightness could be caused by emission from
very small grains or possibly PAHs (Flagey et al. \cite{Flagey2006}). If, as
it seems likely, the emission is restricted to regions of low extinction and
mainly to $\lambda \ga 2\mu$m, it will not cause serious problems for the
column density estimation. If Eq.~\ref{eq:c1} is modified to accommodate the
additional emission component, colour excess data of background stars can
still be used to derive the relation between K-band surface brightness and
column density for higher $A_{\rm V}$. This question is discussed further in
the Appendix~\ref{sect:other}.  Further modifications to Eq.~\ref{eq:c1} will
be needed if the surface brightness of the background sky is not negligible.
At high galactic latitudes the extragalactic infrared background is the main
background component. If completely ignored, it could cause systematic errors
again at the level of 10\% percent (see Appendix~\ref{sect:background}).
However, if the parameters of Eq.~\ref{eq:c1} are determined through
comparison with extinction data, most of this error is already automatically
eliminated.

The scattered light traces the distribution of dust, and the total column
density can be estimated by assuming a constant gas-to-dust ratio. Conversely,
the scattered surface brightness provides a way to study the relative
distribution of gas and dust. In the NIR one can easily reach sub-arcsecond
resolution, which, at the distance of the closest star forming regions
corresponds to only $\sim$100\,AU. Comparison with interferometric
observations of suitable molecular tracers might reveal entirely new
phenomena, such as the clustering of dust grains within turbulent flows
(Padoan et al. \cite{Padoan2006b}).

\section{Conclusions} \label{sect:conclusions}

We have examined the use of scattered NIR surface brightness as a high
resolution tracer of the structure of interstellar clouds. Column densities
are estimated using the J, H, and K bands. In the optically thin case the
observed surface brightness is directly proportional to column density.
In clouds of moderate optical thickness, $A_{\rm V}$ below $\sim
15^{\rm mag}$, the saturation of the NIR surface brightness can be corrected,
and reliable estimation of column densities is still possible. We have studied
how the column density estimation is affected by cloud structure, optical
depth, and dust property variations. In the numerical models the absolute
accuracy of the estimated column densities is better than $\sim$20\%. Large
variations in the radiation field or dust properties can cause large errors,
if not corrected for. However, even in those cases the errors are mostly
systematic, and the relative errors between map pixels are significantly
lower than the systematic errors. 

Sensitive observations of the surface brightness will automatically
provide colour excess data on background stars that can be used as a largely
independent measure of the column density. We showed that by combining the two
methods most of the errors caused by anomalous radiation field or dust
properties can be easily corrected, resulting in a reliable column density map
with significantly higher resolution than is possible with colour excess data
alone. Conversely, the comparison of the two column density estimates can give
important information on the radiation field and dust property
variations within the cloud, and even give hints on its three-dimensional
structure.

\acknowledgements

M.J. and V.-M.P.  acknowledge the support of the Academy of Finland Grants no.
206049 and 107701. P.P. was partially supported by the NASA ATP grant
NNG056601G and the NSF grant AST-0507768.

\appendix

\section{Use of iterative radiative transfer modelling}
\label{sect:detailed_modelling}

The assumption of a uniform radiation field breaks down in optically thick
regions (see Fig.~\ref{fig:simu_errors}b), and even in a homogeneous cloud the
intensity gradients increase the errors of the column density estimates. These
variations are related to the three-dimensional cloud structure and,
therefore, can never be fully eliminated. A rough correction is possible, if
the cloud structure along the line-of-sight can be ignored, and only the
projected column density map is used. Variations in the radiation field can be
estimated with radiative transfer calculations. An initial model cloud is created
according to preliminary estimates of the column densities. In the
line-of-sight the structure is not restricted by observations, and we start by
assuming a Gaussian density profile with FWHM equal to one third of the cloud
size. Radiative transfer calculations are used to produce maps of surface
brightness, and the column densities of the model cloud are corrected
iteratively, using the ratio of the observed and modelled intensities. The
correction is done separately for each sightline corresponding to a pixel in
the maps. 

The correction procedure was first applied to model $F$ with $A_{\rm V}=1.6$
and on model $D$ with $A_{\rm V}=3.2$. Cloud $F$ is very inhomogeneous and,
therefore, a correction without any knowledge of the line-of-sight structure
is expected to be more difficult than for the more homogeneous cloud $D$ (see
Fig.~\ref{fig:av_distributions}). Since the iterative model fitting is
somewhat time-consuming, the model clouds were downsized to 64$^3$ cells. The
relative rms noise of the initial simulated maps was below 1\%. This applies
both to the initial 'observed' surface brightness maps and the simulated maps
produced during the iterations.

\begin{figure} 
\resizebox{\hsize}{!}{\includegraphics{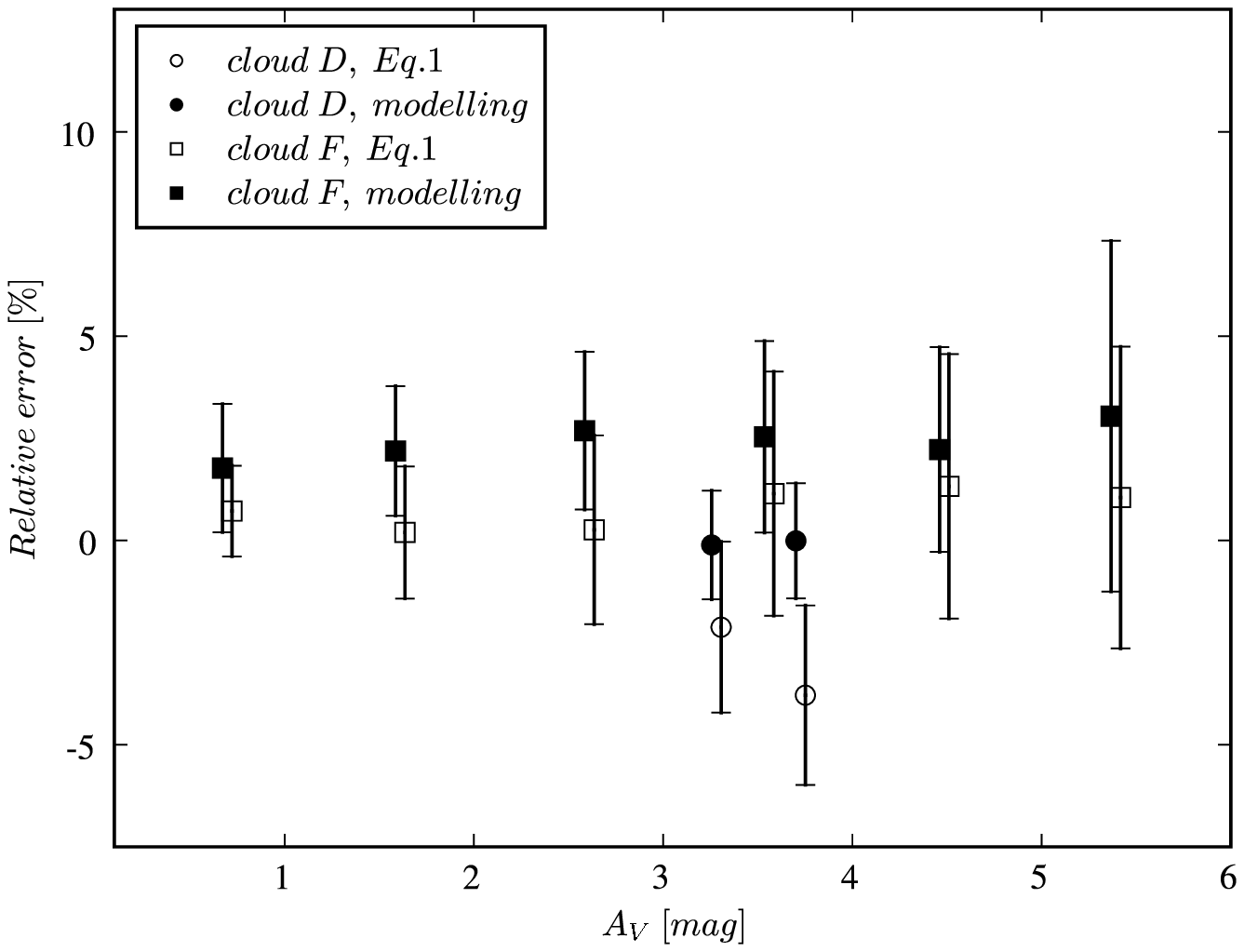}}
\caption{
Comparison of the accuracy of column density estimates obtained with the
analytical method (open symbols) and with detailed radiative transfer
modelling (solid symbols). Results are shown for model $D$ with $<A_{\rm
V}>=3.2^{\rm m}$ and for model $F$ with $<A_{\rm V}>=1.6^{\rm m}$. The plot
shows the bias and the scatter with respect to the true column densities.
} 
\label{fig:opt1}
\end{figure}

In Fig.~\ref{fig:opt1} we compare the resulting column density estimates with
those obtained using Eq.~\ref{eq:c1} together with parameters $a$ and $b$ that
were averages of the values obtained from the six models.  As expected, the
modelling works best for the cloud $D$. In that case estimates are free from
bias and scatter is smaller than obtained with the default method. 
In the latter method, if we used exactly correct parameter values in Eq.~\ref{eq:c1},
the bias would be removed but the scatter would remain unchanged. For model
$F$ the accuracy is rather similar for the two methods. The iterative
modelling shows a systematic error 2-3\%, probably as a consequence of the
assumption of a smooth line-of-sight density distribution.

\begin{figure} 
\resizebox{\hsize}{!}{\includegraphics{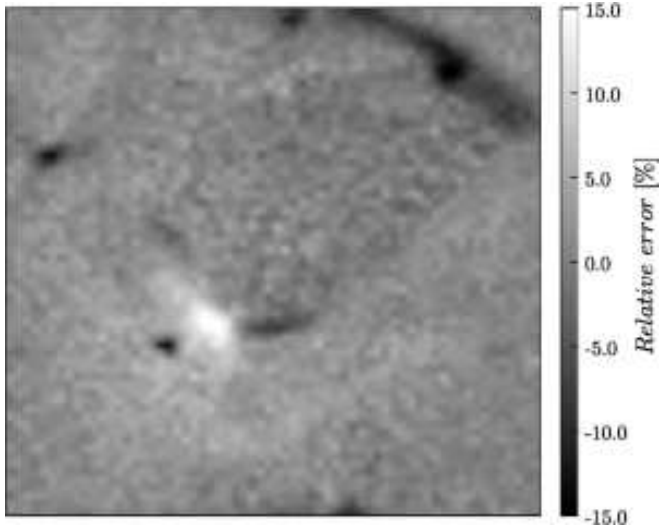}}
\caption{
Relative errors in the column densities obtained for model $C$,
$<A_{\rm V}>=1.6^{\rm m}$, with iterative radiative transfer modelling. The
model assumed Gaussian density distribution along the line-of-sight.
} 
\label{fig:opt3}
\end{figure}

\begin{figure} 
\resizebox{\hsize}{!}{\includegraphics{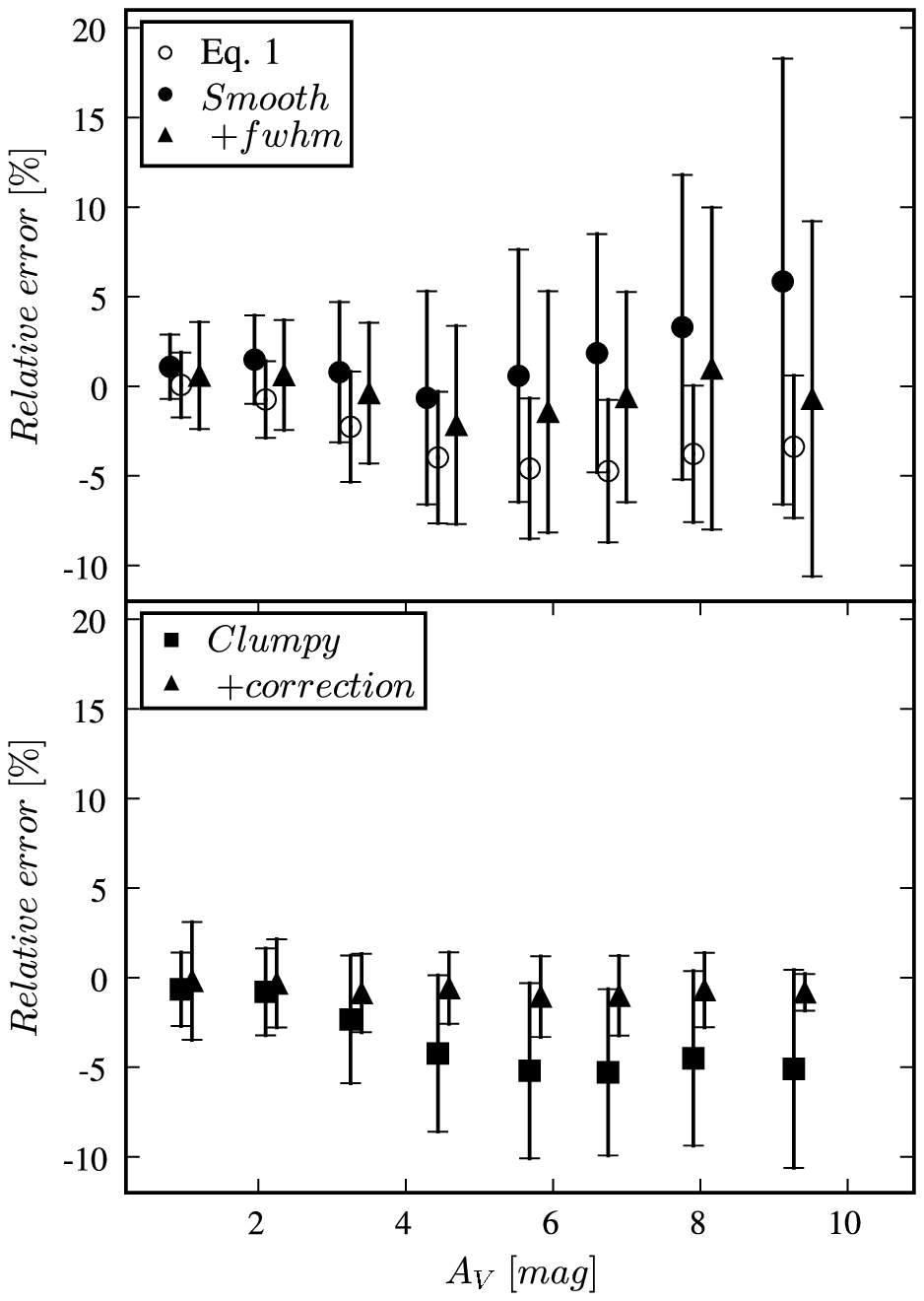}} 
\caption{
Accuracy of column density estimates in the case of model $C$, $<A_{\rm
V}>=1.6^{\rm m}$, when three different methods are used. {\em Upper frame}:
Original method using Eq.~\ref{eq:c1} (open circles), iterative modelling with
Gaussian line-of-sight density distribution (solid circles), and iterative 
modelling with Gaussian line-of-sight density distribution with width taken as a
free parameter for each sightline (solid triangles). {\em Lower frame}: Results 
from iterative modelling with clumpy density distribution (squares), and the same 
after a simple correction based on the error in the predicted J-band intensity (triangles).
} 
\label{fig:opt4} 
\end{figure}

The use of Eq.~\ref{eq:c1} includes the effects of a clumpy cloud structure
since the fit parameters are obtained from inhomogeneous models. On the other
hand, it ignores any spatial variation on the plane of the sky. The iterative
modelling worked differently, fitting the spatial variations but using a
simplistic description for the line-of-sight density structure. Remaining
errors in the column densities are a direct consequence of the true cloud
structure differing from the smooth distribution used in the modelling. This
is more clear in the case of model $C$ for which error maps are shown in
Fig.~\ref{fig:opt3}. In this case the iterative modelling results in a much
larger scatter (see Fig.~\ref{fig:opt4}). The maps contain intriguing
features, where the column density is either systematically overestimated or
underestimated. The same regions can be easily recognized without knowledge of
the true column density because, when column density is underestimated, the
predicted J-band intensity from the iterative modelling is relatively higher
and the K-band intensity lower than in the actual model $C$. This is caused by
the fact that, because of the smaller saturation, the K-band intensity is more
sensitive to changes in the column density. Comparison with
Fig.~\ref{fig:colden_maps}b shows that the deviations are not necessarily
associated with the largest column densities. There could be several
explanations for such deviations. First, a dense region close to the cloud
surface could produce more surface brightness (especially in the J-band) and
could result in an overestimation of the column density. Examination of the
three-dimensional density distribution does not support this interpretation.
The only region with clearly overestimated column density values is located
close to the centre of cloud $C$. Similarly, the strongest negative features
are caused by filaments that are relatively close to the cloud surface. The shape
of the condensations may also play a role, the surface brightness being
slightly higher when the material is distributed along a longer sightline.
This agrees better with the structures seen in the cloud $C$. When the
line-of-sight FWHM of the density distribution was used as a further free
parameter, the previous fit results did indeed improve significantly (see
Fig.~\ref{fig:opt4}a).

The scattered surface brightness might also carry some information about the
relative positions of the dense regions. If scattering occurs mostly in the
forward direction, the surface brightness samples the average radiation field
behind the clumps. In that case the clumps closest to the observer would have
lower ratios between the surface brightness and the column density. However,
this seems to be a very small effect. When model $C$ is observed from the
opposite direction, the resulting error maps do, within noise, exactly
correspond to those shown in Fig.~\ref{fig:opt3}. Finally, the deviations
might be caused by the general lack of homogeneity, which would affect the
way radiation can penetrate the cloud. This seems to be, at least partially,
an explanation for the underestimated column densities in the cloud centre.
This was tested in a qualitative way by using the density distribution of
cloud $D$ in the fitting of model $C$ surface brightnesses. The initial
line-of-sight densities were random (models $D$ and $C$ are completely
independent runs), and the column densities corresponding to each map pixel
were again optimized. The result is shown in Fig.~\ref{fig:opt4}b. The scatter
is now rather similar to what was obtained by using Eq.~\ref{eq:c1}. The use
of an inhomogeneous medium eliminates overestimated column densities in the
central region. However, the negative features seen in Fig.~\ref{fig:opt3} do
remain. These might be caused by the compactness of the emitting regions, as
discussed above. Because the deviating regions are visible as excess in the
simulated J-band maps, it should be possible to make a correction for those,
for example by using the line-of-sight size of the clumps as a further free
parameter. Figure~\ref{fig:opt4}b shows results after a simple correction where
the column density estimates are increased by the ratio of modelled and true
J-band intensities. This is just to illustrate that further improvements are
possible.

When optical depth becomes significant, an increase in the column density has
gradually less and less effect on the observed intensity. The saturation
depends on the wavelength, and changing intensity ratios between different
bands carry information on the column density. In the modelling this can be
used to break the degeneracy between radiation field and column density. This
was tested with the model $C$ and repeating the optimization with the scaling
of the external radiation field as free parameters. When observational errors
were included, the correct intensity level could still be recovered with
accuracy better than 10\%.  This shows that even if the intensity of the
radiation field were not a priori known, the final uncertainty would not be
significantly larger than shown in Fig.~\ref{fig:opt4}.

\section{Validity of the basic assumptions} \label{sect:validity}

In this section we discuss possible violations of the basic assumptions behind
the use of Eq.~\ref{eq:c1} and examine their effect on the accuracy of the
column density estimates.

\subsection{Anisotropy of the radiation field}

The assumption of a uniform radiation field is violated if (1) the background
radiation is anisotropic, (2) the cloud is sufficiently optically thick to
cause significant attenuation in its interior, or (3) the cloud contains
strong internal radiation sources. The anisotropy of the radiation field was
not studied explicitly. However, it is clear that it becomes important only if
both conditions (1) and (2) are true. A gradient in the intensity of the
incoming radiation causes a corresponding gradient in the column density
estimates. In the models the attenuation of the radiation field caused errors
only at 10\% level between cloud edge and cloud centre. The sky brightness
depends on the Galactic coordinates, so that the ISRF is never fully
isotropic. The clouds are still illuminated from all sides, and the Galactic
plane fills a large solid angle. Therefore, also the difference between cloud
edges should exist only at a level below $\sim$10\%. Although shadowing by
dense regions can cause variations even at relatively small scales, this can
still be identified with the help of an extinction map. Once the gradient has
been identified or the source of the anisotropy is known, it can be corrected,
either by modelling the radiation field or by varying the fit parameters of
Eq.~\ref{eq:c1} as the function of position. Direct re-scaling of the derived
column density estimates does not work as well, because of the non-linearity
of Eq.~\ref{eq:c1}. In the near-infrared the scattering takes place mostly in
the forward direction. This means that the observed surface brightness depends
more on the sky brightness behind the cloud than in other directions, and the
scaling between surface brightness and column density changes as a function of
Galactic coordinates. Of course, this can be corrected by taking into account
the true sky brightness distribution available from DIRBE observations. In the
first approximation the effect is eliminated if the method is re-calibrated
using background stars. There is a small difference between optically thin and
optically thick regions, because, in the latter, multiple scattering tends to
make the field more isotropic. However, when the surface brightness is used
for the column density estimation, radiation can be only moderately optically
thick, and the same correction factor should be valid for the whole map.

The radiation field is always attenuated within the cloud. In a homogeneous
cloud the gradient is systematic and can be estimated by modelling (e.g.,
Fig.~\ref{fig:opt1}). In practice the radiation field depends on the unknown
3D cloud structure. Although an exact correction is not possible, our model
calculations showed that, by taking into account some general properties of
the density field, the accuracy of the column density estimates can be
improved (e.g., Fig.~\ref{fig:opt4}). Our results apply to clouds with $A_{\rm
V}$ below 20$^{\rm m}$. This limit corresponds to a K-band optical depth of
$\sim 1.8$. For regions with $\tau_{\rm K}>>1$ the scattered NIR light can not
be used as a tracer of the total column density. As the optical depth exceeds
unity, the surface brightness saturates and, if the region is sufficiently
large, finally decreases towards the column density maximum. The resulting
limb-brightening serves only as an indicator of the radius at which the medium
becomes optically thick for the observed radiation.

\subsection{Internal sources} \label{sect:internal}

Internal radiation sources can cause local increase in the surface brightness
and decrease the accuracy of the column density estimates in active clouds. If
a source is deeply embedded its effect on the surface brightness will be
limited to a very small area or it can be readily identified, either
directly from surface brightness maps or from a comparison with an extinction
map.  If the source is outside dense regions, its effect may be more subtle,
as was seen in Fig.~\ref{fig:sources}a. A source can illuminate distant
filaments while, in projection, nearby regions are apparently unaffected.  As
an example, we consider the Chamaeleon I cloud where some one hundred embedded
sources have been identified (G\'omez \& Kenyon \cite{Gomez2001}). The cloud
is located at the distance of 160\,pc, and on the sky the sources are at mean
distance of a few arc minutes from each other. The brightest source has
$m_{\rm K}=10.2^{\rm m}$, five sources are brighter than 12$^{\rm m}$ but most
sources are fainter than 13$^{\rm m}$. Even the strongest source is fainter
than the weakest source in Fig.~\ref{fig:sources}. If this source were located
in a dense region, it could be visible in the surface brightness of scattered
light, but the effect would be limited to a very small area. Similarly, the
contribution of 13$^{\rm m}$ sources located at one arc minute intervals
remains almost two orders of magnitude below the ISRF intensity. If the source
is really embedded inside the cloud, the associated elevated NIR dust {\em
emission} should be limited to an even smaller area. Therefore, such embedded
sources do not seriously affect the accuracy of the column density estimates.

If cloud contains very strong sources, only in very simple cases can their
effect be fully modelled. However, comparison of extinction and surface
brightness maps gives a direct way to estimate the effect of internal
radiation sources and, at the same time, gives clues about the 3D structure of
the object. When the two column density estimates agree, the surface
brightness method can be safely used to extend the column density estimation
to smaller spatial scales. Close to a source, the surface brightness maps does
contain important information about the cloud structure, although weighted
according to varying radiation field. In practice, the discrepancies between
the two column density estimates can be caused also by problems in the
extinction map, for example, if the colour excess data are contaminated by
foreground stars or background star cluster or galaxies.

\subsection{Dust properties} \label{sect:dust_properties}

Dense regions may be problematic, not only because of the attenuation of the
radiation field but also because of changes in dust properties. In
Sect.~\ref{sect:dust} we examined how the results are affected if dust
properties do not agree with the model parameters of Eq.~\ref{eq:c1}. Below
$A_{\rm V}\sim 10^{\rm m}$ the effect was only $\sim$20\%. This is almost
entirely a systematic effect and does not affect the relative accuracy between
regions of different column density. Only above $A_{\rm V}\sim 10^{\rm m}$
does the non-linearity of Eq.~\ref{eq:c1} produce significant $A_{\rm
V}-$dependence. In this case no attempt was made to correct the values of the
erroneous model parameters. In Sect.~\ref{sect:extinction} we examined a more
realistic case where dust properties changed according to the local density,
from $R_{\rm V}=3.1$ in low density regions to $R_{\rm V}=5.5$ at the highest
densities. This time a comparison with low-resolution extinction map was used
to re-estimate the parameters of Eq.~\ref{eq:c1}. The result was a well
calibrated column density map that had high spatial resolution and was was
morphologically very accurate. One set of global parameters was used in
Eq.~\ref{eq:c1}. Although this did work reasonably well for both the dense and
the diffuse regions, the use of spatially varying parameters might still bring
some improvement.

Dust properties affect the ratio between scattered surface brightness and
extinction (see Sect.~\ref{sect:extinction}). When grains become larger the
albedo increases and scatterings are oriented more in the forward direction.
This affects the transport of radiation during the first phase (see
Sect.~\ref{sect:method} and Fig.~\ref{fig:scheme}) and, for a given
extinction, makes the field more uniform within the cloud.  For clouds with
$A_{\rm V}\sim 10^{\rm m}$ the variations on the plane of the sky are small,
and changes in the albedo and scattering function can only have a minor
effect. The main effect comes from the fact that the observed surface
brightness is directly scaled by the albedo at the position of the last
scattering. During the second phase the radiative transfer equation is similar
as for the extinction of the background stars and depends on the extinction
cross section. The net effect is that the albedo acts as a multiplicative
factor that scales the surface brightness relative to the extinction.
Conversely, the ratio of surface brightness and extinction could be used to
trace variations in the average grain size. However, as shown by
Fig.~\ref{fig:bicomponent}, the effect may be small and its detection and
separation from all the other possible effects is not straightforward.

\subsection{Other sources contributing to the NIR surface brightness} \label{sect:other}

We have assumed that the observed surface brightness is due to
dust scattering inside the cloud. If there are other significant sources,
their contribution must be determined and removed or the column density
estimates are correspondingly in error. We discuss three potential factors
affecting the NIR surface brightness: Thermal dust emission, emission lines,
and the role of the NIR background.

\subsubsection{Thermal dust emission} \label{sect:thermal_emission}

At optical wavelengths scattering dominates over the thermal emission from
cold dust particles. In the mid-infrared the situation is already reversed and
emission from transiently heated small particles and large molecules is the
dominant component. In the mid-infrared the strong emission bands are
generally attributed to PAH molecules. However, there is also a broad
continuum that extends to the NIR, although significantly diminished. We
estimate the thermal dust emission based on the model of Li \& Draine
(\cite{Li2001}). According to their Fig.~8, the expected dust emission in the
K band is $\lambda I_{\lambda} \sim 2\times 10^{-27}$\,erg/s/sr/H. Excluding
all absorption, the expected intensity for dust column with $A_{\rm V}=1.6$ is
$\sim 4.5 \times 10^{-20}$\,erg/s/cm$^2$/sr/Hz. For similar $A_{\rm V}$ the
surface brightness of scattered radiation was in our calculations $\sim 2
\times 10^{-19}$\,erg/s/cm$^2$/sr/Hz, the number including some reduction
caused by absorption. Therefore, in an optically thin case the emission could
contribute some 20\% to the total surface brightness. However, while the
scattering involves here only one NIR wavelength, dust emission depends on the
heating that takes place through shorter wavelengths. This is especially true
for the NIR emission that requires grain temperatures of the order of 1000\,K
and, consequently, heating by energetic photons. At $A_{\rm V}=1^{\rm m}$ the
attenuation is in the U-band a factor of $\sim$5 and even the V-band close to
a factor of three. Therefore, dust emission is restricted to cloud surfaces,
and, even in translucent clouds, its contribution to the total surface
brightness is only a few per cents. In more optically thick clouds the ratio
between scattered and emitted intensity increases, until also the scattered
K-band intensity saturates after $A_{\rm V}\sim 10^{\rm m}$. In the H-band the
expected dust emission is only half of the value in the K-band. Moreover,
there is evidence that the abundance of small grains decreases rapidly in
dense clouds (e.g., Stepnik et al. \cite{stepnik2003}), further reducing the
relative amount of NIR emission expected from optically thick clouds.

Compared with the continuum, the contribution of possible PAH lines and other
features is small (e.g., Gordon et al. \cite{Gordon2000}, Mattioda et al.
\cite{Mattioda2005b}b). In the Li \& Draine model small graphite grains are
responsible for most of the NIR continuum. For {\em ionized} PAHs the NIR
absorption cross sections may be significantly higher than assumed by that
model (see Mattioda et al. \cite{Mattioda2005a}a, \cite{Mattioda2005b}b). 
This may affect the interpretation of the source of near-infrared continuum
but, because the modelled small grain contribution was well above the PAH
continuum, the effect on total predicted emission would be much smaller.
Furthermore, within most of the volume of our clouds the radiation field is
both softer and weaker than the normal ISRF, thus reducing the degree of PAH
ionization.  On the other hand, Sellgren et al. (\cite{Sellgren1996} and
references therein) observed in several optical reflection nebulae significant
NIR dust emission that was similarly attributed to either very small,
transiently heated dust grains or PAH-type molecules. In the K band the
emission exceeded the predicted scattering when the temperature of the central
star was larger than $\sim 6000$\,K. One can make a conservative assumption
that the ratio of the emitted and scattered radiation follows the ratio of
incoming K-band and V-band intensities. For a 6000\,K black body this ratio is
1.66, while in the case of the ISRF (Mathis et al. \cite{Mathis83}) the ratio
is only 0.42, i.e., lower by a factor of four. In reality the reduction can be
much larger, because the heating is likely to require large flux of
UV-photons, because of the strong attenuation of the optical-UV field in
optically thick clouds, and because the observed NIR emission may be related
to a difference in dust properties. Therefore, the results of Sellgren et al.
are not in contradiction with our assumption that in a cloud illuminated by
normal ISRF the scattering stands for most of the observed surface brightness.

Further evidence for NIR dust emission comes from DIRBE observations. Bernard
et al. (\cite{Bernard1994}) presented a spectrum for the average cold diffuse
medium at $|b|=5\deg$. After subtraction of zodiacal light and stellar
emission, the remaining signal was 0.023\,MJy/sr at 3.5$\mu$m and 6.7\,MJy/sr
at 100$\mu$m, corresponding to $A_{\rm V}\sim$0.5.  Only a minor fraction
of the 3.5$\mu$m signal can be due to scattered light. If one assumes that the
3.5\,$\mu$m value is caused entirely by emission and further assumes a flat
spectrum, $\nu I_{\nu}$=constant, the predicted K-band emission would be 
1.5$\times10^{-19}$\,erg/s/cm$^{-2}$/sr/Hz. For this low $A_{\rm V}$ the
value is higher than the expected intensity of the scattered light, while in
the J-band the emission would drop to $\sim$10\% of the scattered intensity.
Arendt et al. (\cite{Arendt1998}) correlated different DIRBE bands against the
100\,$\mu$m band at $|b|<30\deg$. The correlation coefficient between
3.5$\mu$m and 100$\mu$m intensities was, 1.8$\times 10^{-3}$, i.e., lower by
almost a factor of two. 
The values might still be overestimated, if the DIRBE values contained
additional signal from incompletely removed stars or from scattering, or were
significantly affected by the presence of the 3.3$\mu$m PAH feature.

Flagey et al. \cite{Flagey2006} have recently studied NIR emission using
ISO and Spitzer data and modelled emission seen at diffuse sightlines. Based
on their model the extrapolated value at 2$\mu$m is $\sim$0.03\,MJy\,
sr$^{-1}$ for $N_{\rm H}=10^{21}$\,cm$^{-2}$ (see their Fig. 9). At $A_{\rm
V}=$0.5$^{\rm m}$ this would correspond to
1.9$\times10^{-19}$\,erg/s/cm$^{-2}$/sr/Hz, i.e., a slightly higher intensity
than the previous Bernard et al. (\cite{Bernard1994}) value. The NIR continuum
was modelled with a 1100\,K gray body which is typical of reflection nebulae.
A lower colour temperature of the continuum would very rapidly reduce the
predicted K-band intensity. The model of Flagey et al. \cite{Flagey2006}
applied to diffuse medium in the inner Galaxy where, according to their own
estimate, the radiation field has intensity three times the local ISRF. The
emission should be correspondingly smaller in local clouds.
A critical question is, at what $A_{\rm V}$ the UV-field has
attenuated so much that the scattering dominates also the K-band surface
brightness. 

The NIR dust emission remains uncertain because of the lack of observations
and because of the significant uncertainties in the modelled NIR emission (see
also Zubko et al. \cite{Zubko2004}). There is observational evidence that at
higher column densities the NIR surface brightness is clearly dominated by
scattering (e.g., Lehtinen \& Mattila \cite{Lehtinen1996}; Foster \& Goodman
\cite{Foster2006}). The DIRBE results are consistent with those observations,
if the dust emission comes mainly from diffuse material. In dense clouds, the
emission would be restricted to a narrow surface layer and it would be
distributed rather evenly over the whole cloud. Part of the signal may be
removed in on-off observations, and the remainder would not be strongly
correlated with the column density. Furthermore, if the parameters of
Eq.~\ref{eq:c1} are determined from observations, e.g., from comparison with
extinction data, the contribution of dust emission is already taken into
account. The functional form of Eq.~\ref{eq:c1} can accommodate some
contribution from emission. If emission is strong, an additional offset term
is required. If the K-band turns out to contain significant emission, its use
can be restricted to opaque regions, where dust emission is low. Indeed, the
K-band data is essential only in very dense regions, where the shorter
wavelength data suffer from significant saturation. Furthermore, the K-band
observations are relatively more time-consuming, and at low column densities
the S/N-ratio tends to be worse than for the other bands. Conversely, deep NIR
observations will enable mapping of the relative intensity of the dust
emission and scattering.

\subsubsection{Emission lines} \label{sect:emission_lines}

The NIR bands can contain line emission from smaller molecules and ions. In
cold clouds the excitation is far too low to produce strong NIR lines, e.g.,
from CO and H$_2$ molecules. Significant line emission can occur either in
photon dominated regions (PDRs) on the cloud surface or in shocks triggered,
for example, by the outflows from young stellar objects (YSOs) within the
cloud.  The presence of strong PDR regions would also contradict the
assumption of an isotropic background. Foster \& Goodman {\cite{Foster2006}}
discussed the possible contribution of H$_2$ lines based on the models of
Black \& Dishoeck (\cite{Black1987}). They noted that the observed J- and
H-band intensities showed no indication of line emission. Indeed, in the
models of Black \& Dishoeck, assuming a cloud illuminated with normal ISRF,
the effect of H$_2$ should remain insignificant.

YSOs are commonly associated with molecular outflows that are driven by fast
jets. Along its path the jets cause shocks, Herbig-Haro objects, that are
visible in the NIR, especially through many H$_2$ lines.  The strongest line,
the H$_2$ 1-0 S(1) line at 2.12$\mu$m, falls within the K-band filter. The J
and H bands have similarly some contribution from, e.g., the excited [FeII]
lines at 1.24 and 1.64$\mu$m (Reipurth \cite{Reipurth1999}). Therefore, along
the jet all NIR bands may be contaminated by line emission. However, if
Herbig-Haro objects are observed, they are usually clearly aligned along the
jets, which must be masked out from subsequent analysis. If no outflow is
detected, for example, if even the brightest knots fall below the detection
limit, the jet will clearly have no effects on the column density estimates.
At the end of the outflow then bowshocks may also cause enhanced NIR surface
brightness. Bowshocks can be much larger than the Herbig-Haro regions, and may
be located very far from the driving source and the visible part of the jet.
The shape of a bowshock depends on the density structure of the ambient
material that is interacting with outflow. Therefore, a purely morphological
identification of bowshocks may be difficult. Strong bowshocks can be
recognized based on anomalous intensity ratios, while faint ones may go
unnoticed and cause some errors in the column density estimates.

\subsubsection{Background radiation} \label{sect:background}

In on-off measurements one measures the difference between the signal from the
source and the background. If the intensity of the diffuse background, $S_{\rm
bg}$ is large behind the source without significantly affecting the total
amount of radiation entering the cloud, the observed signal will decrease by
an amount of $S_{\rm bg}(1-e^{-\tau})$, where $\tau$ is the optical depth of
extinction for a sightline through the cloud. As long as this term is smaller
than the surface brightness due to scattering, the dependence between column
density and surface brightness will only become more shallow. As an empirical
description of this dependence Eq.~\ref{eq:c1} should remain approximately
valid, provided that the parameters of the equation take the background into
account. Re-calibration using the colour excess data would automatically
accomplish this. If $S_{\rm bg}$ is equal to the surface brightness of
scattered light, column densities could no longer be estimated, because the
observed signal would be zero. If $S_{\rm bg}$ is larger and sufficiently
constant, the cloud structure could, of course, be traced by the absorption.

We will first consider the cosmological IR background (CIRB), $S_{\rm CIRB}$.
If the absorption due to the Galaxy is ignored, this is an isotropic
component. If one considers optically thin clouds or a perfectly spherical
cloud, the {\em scattered} CIRB photons have no effect on the observed surface
brightness: The distribution of outcoming photons is also isotropic. If the
cloud is optically thick and inhomogeneous, the radiation field is no longer
isotropic inside the cloud. Therefore, the amount of radiation scattered
away from one line-of-sight is no longer equal to the amount of radiation
scattered from other directions into that direction. However, this imbalance is
only a small fraction of the scattered CIRB intensity and this variation in
the surface brightness of the scattered CIRB photons can be safely ignored.
Because of the {\em absorbed} radiation the observed intensity decreases by
$S_{\rm bg}(1-e^{-\tau})$, the term $\tau$ being the optical depth for
absorption. In the optically thin case the scattering can be ignored altogether,
because each scattered photons escapes the cloud and the surface brightness of
the cloud is identical with the background sky. The depression in the surface
brightness of optically thicker clouds has been calculated by Mattila
(\cite{Mattila1976}, table A2). When the optical depth is below 4 and the
contribution from scattering is ignored, the relative error in our estimate of
the change in the surface brightness is less than $\sim$15\%. 
This is sufficient accuracy for our purposes.

\begin{figure} 
\resizebox{\hsize}{!}{\includegraphics{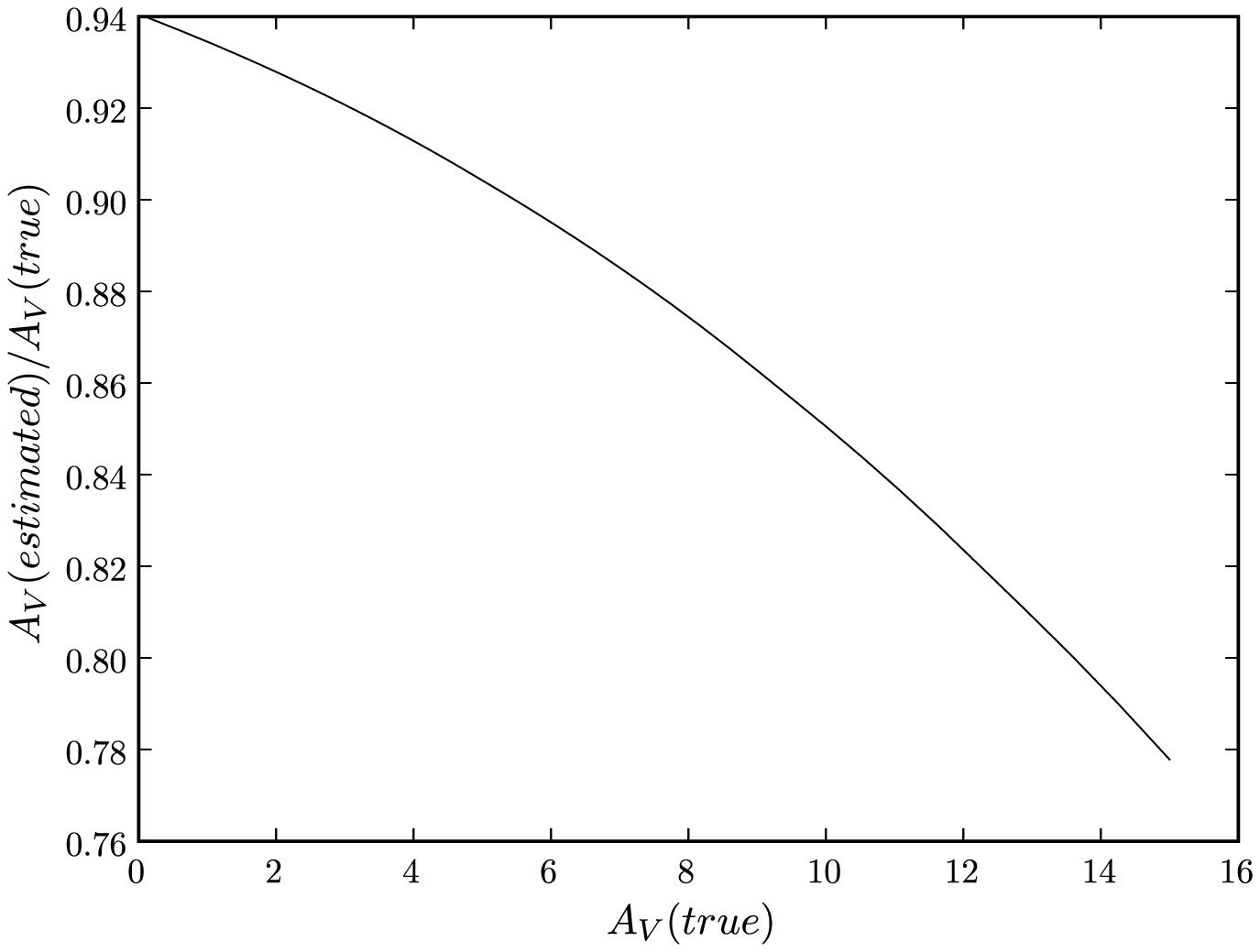}} 
\caption{
Ratio of column density estimates and true column density when the analysis 
ignores an existing background surface brightness. In the plot the column
densities are derived based on K-band data and using the average parameters
obtained from the numerical models. The background intensity is $1.64 \times
10^{-19}$\,erg/s/cm$^2$/Hz/sr (see text).
} 
\label{fig:ebl} 
\end{figure}

The tentative K-band CIRB detection of Gorjian, Wright, and Chary
(\cite{Gorjian2000}) was $S_{\rm bg}=1.64 \times
10^{-19}$\,erg/s/cm$^2$/Hz/sr. This high value corresponds almost to the mean
surface brightness of scattered radiation in our $<A_{\rm V}>=1.6^{\rm m}$
models. In on-off observations the addition of this background would decrease
the measured K-band surface brightness excess roughly by $S_{\rm bg}
(1-e^{-0.049\times A_{\rm V}})$, the factor 0.049 being the ratio of K-band
absorption and $A_{\rm V}$. The decreased surface brightness leads to an
underestimation of the column densities. The magnitude of this effect is shown
in Fig.~\ref{fig:ebl}, assuming that the parameters of Eq.~\ref{eq:c1} are
determined {\em without} taking the background into account and only K-band
data are used.  If the background level were known or the parameters were
corrected using the extinction data, the error could be almost completely
removed. Therefore, in practice it has little effect on the accuracy of the
column density estimates. The CIRB intensity used above is very uncertain, and
the true value may turn out to be significantly lower (e.g., Dwek, Arendt,
Krennrich \cite{Dwek2005}; Aharonian et al. \cite{Aharonian2006}). On the
other hand, the effect of the CIRB is relatively higher in the J- and H-bands,
because of the higher optical depth and possibly higher CIRB intensity (Hauser
\& Dwek \cite{Hauser2001}).

In the modelling we have assumed that the cloud is illuminated by the normal
ISRF. The diffuse background was not included in the calculated surface
brightness nor was a background subtraction included in the subsequent
analysis. If the ISRF is caused by discrete point sources, no background
subtraction is necessary, because the surface brightness can be determined on
the empty sky between sources.  However, if the ISRF contains a significant
diffuse component, the observed values will again be reduced by $S_{\rm
bg}^{diffuse}(1-e^{-\tau})$, $\tau$ being the optical depth for absorption,
and qualitatively the effect is the same as in Fig.~\ref{fig:ebl}. If the
background intensity is approximately isotropic, the main effect comes from
absorption, and the scattering can again be ignored. Diffuse background could
consist of unresolved stars or of photons that are scattered from background
clouds. In the latter case the mapping of the foreground cloud may be
impossible, because its contribution to the observed surface brightness cannot
be separated from the background. The effect of faint stars depends on the
direction on the sky and the resolution and sensitivity of the observations. A
crude estimate can be made assuming a slope of 0.35 in the ($mag$, $log_{10}
N$)-relation (see Sect.~\ref{sect:extinction}). The relation is shallow enough
that, in the total intensity, the contribution of a magnitude bin decreases
with increasing magnitude. For observations discussed in
Sect.~\ref{sect:accuracy} the limiting K-band magnitude is far fainter than
20$^{\rm m}$. If one assumes that the contribution of stars brighter than
20$^{\rm m}$ can be separated from the diffuse background and stars are
integrated down to 40$^{\rm m}$, the contribution of the unresolved stars
remains below 10\% of the total surface brightness. The net effect on the
column density estimates is again some fraction of this. For a given direction
on the sky, more accurate estimates and possible correction can be made using
Galactic models of stellar distribution (e.g., Cohen \cite{Cohen1994}).
However, our estimate is rather conservative, and, in most cases, the effect
of unresolved stars can be ignored. The Galactic plane may be an important
exception, because of the anisotropy of the radiation field, and the source
confusion which makes the separation of the diffuse background difficult. 

In the NIR the zodiacal light is by far the strongest diffuse component but,
because it has a smooth spatial distribution and it resides between the observer
and the studied cloud, it has little effect on the column density estimation.
Some problems can still be caused by surface brightness gradients across large
maps close to the ecliptic plane, or, in rare cases, comet trails that
introduce small scale structure in the surface brightness (e.g., Kelsall et
al. \cite{Kelsall1998}; Nakamura et al. \cite{Nakamura2000}).

\end{document}